\documentclass[pdflatex,sn-mathphys-ay]{sn-jnl}

\usepackage{graphicx}%
\usepackage{multirow}%
\usepackage{amsmath,amssymb,amsfonts}%
\usepackage{dsfont}%
\usepackage{amsthm}%
\usepackage{mathrsfs}%
\usepackage[title]{appendix}%
\usepackage{xcolor}%
\usepackage{textcomp}%
\usepackage{manyfoot}%
\usepackage{booktabs}%
\usepackage{algorithm}%
\usepackage{algorithmicx}%
\usepackage{algpseudocode}%
\usepackage{listings}%
\usepackage{subcaption}%
\usepackage{placeins}

\theoremstyle{thmstyleone}%

\theoremstyle{thmstyletwo}%

\theoremstyle{thmstylethree}%

\usepackage{tikz}
\renewcommand{\orcidlogo}{%
  \begin{tikzpicture}[baseline=-0.15em]
    \fill[color={rgb,255:red,166;green,206;blue,57}] (0,0) circle (0.38em);
    \node[white,font=\bfseries\sffamily\tiny] at (0,0) {iD};
  \end{tikzpicture}%
}

\begin{document}

\title[Reifying Research Logic]{Reifying Research Logic: AI-Assisted Workflow Construction and Incremental Refinement for Quantitative Syntax}


\author[1]{\fnm{He} \sur{Wang}\orcid{https://orcid.org/0009-0009-1274-8319}}
\author[1]{\fnm{Jingbo} \sur{Chen}\orcid{https://orcid.org/0009-0005-5177-5513}}
\author[1]{\fnm{Yuqiao} \sur{Lai}\orcid{https://orcid.org/0009-0006-9937-1232}}
\author[1]{\fnm{Nan} \sur{Yang}\orcid{https://orcid.org/0000-0001-5259-3512}}
\author[1]{\fnm{Hanwen} \sur{Zhang}\orcid{https://orcid.org/0009-0005-5212-5619}}
\author*[1]{\fnm{Wei} \sur{Yuan}\orcid{https://orcid.org/0009-0004-2498-9447}}\email{yuanwei@nudt.edu.cn}

\affil[1]{\orgdiv{College of International Studies}, \orgname{National University of Defense Technology}, \orgaddress{\city{Nanjing}, \state{Jiangsu}, \postcode{210000}, \country{China}}}


\abstract{Quantitative language research often depends on long chains of computational steps, yet the logic connecting those steps usually remains buried in scripts. This makes analyses harder to inspect, share, and revise than they need to be. Focusing on quantitative syntax, we present QLWF, a visual workflow platform that turns natural-language research descriptions into executable workflows through an AI-assisted five-stage pipeline. In this setting, reification makes the research logic visible as a workflow, while formalization gives that workflow deterministic execution semantics. The language model is used only during construction. Execution is handled by a fixed node library and engine, which keeps the resulting workflows reproducible. QLWF also supports incremental refinement, so saved workflows can be revised by changing only the parts that need to change rather than being rebuilt from scratch. To evaluate the approach, we build a 64-task benchmark called QL-Bench from the quantitative-syntax literature. Across three runs, QLWF produces structurally valid and executable workflows for every task and reaches a mean output-plausibility rate of 98.4\%, well above the prompt-based baselines. On a separate 12-task lifecycle benchmark, this refinement process succeeds in every case and uses roughly one-third of the tokens required by full regeneration. The paper also releases the node library, benchmark, workflow templates, and platform as reusable resources for quantitative-syntax research.}


\maketitle


\section{Introduction}\label{sec:introduction}

Quantitative approaches to language research increasingly rely on multi-step computational procedures. These procedures typically span corpus preprocessing, metric extraction, and statistical modelling. In quantitative syntax alone, a single study may involve computing dependency distances from parsed treebanks \cite{liu2008}, fitting distributional models such as Zipf's law to frequency data \cite{zipf1949}, and comparing results across typologically diverse corpora \cite{nivre2020}. Each of these steps requires specific tools, data formats, and parameter configurations. As the number of analytical steps grows, keeping the overall research logic inspectable, reproducible, and open to revision becomes correspondingly harder.

In current practice, however, the analytical procedures behind such studies are rarely made explicit as structured objects. Researchers typically write their analyses as ad-hoc scripts, where the dependencies among processing steps, the data flow between tools, and the choice of parameters all stay buried in the code. The overall research logic is therefore not available in any form that can be inspected independently of the implementation. This is the \textit{representation gap}. At the same time, the individual tools involved (parsers, statistical packages, visualisation libraries) operate in isolation. Platforms such as GATE \cite{cunningham2002}, WebLicht \cite{hinrichs2010}, and TextFlows \cite{perovsek2016} provide graphical interfaces for chaining NLP components, but the analytical logic that connects these components into a coherent research procedure remains hard to capture or reuse as a self-contained object. This is a second problem, the \textit{accessibility gap}, and it shows up most in cross-group collaboration and in later reuse. Once the two gaps compound, the analytical logic of a quantitative study is hard to inspect as a whole, hard to share across groups, and harder still to revise as analytical requirements evolve. It is still not treated as a reproducible research artefact, even though it arguably counts as a language resource in its own right \cite{goble2020}.

One natural response to these gaps is to organise analytical procedures as \textit{workflows}: graph-structured representations in which each processing step is an explicit node and each data dependency is a visible edge \cite{crusoe2022}. Recent work on AI-assisted workflow generation has been moving in this direction. However, the resulting pipelines are often \textit{agent workflows}, where a large language model decides the execution path at runtime. These workflows are genuinely flexible, but the price is losing determinism and reproducibility \cite{zhang2025,tan2025}. For empirical language research, where conclusions must trace back to fixed analytical steps, a different balance is needed. Our approach separates the two: AI assists with construction, while execution itself follows a deterministic, pre-built data flow. Once externalised this way, the research logic can be inspected, shared, and re-run with consistent results. The paper therefore takes the workflow itself as the central object of study. It is a \textit{reified} form of the researcher's analytical intent, human-readable as a visual node graph and machine-executable as a deterministic data pipeline. The researcher owns the final interpretation of the workflow. AI assistance enters only at the construction stage.

Our starting point is \textit{quantitative syntax}, the study of syntactic structure through numerical measures such as dependency distance, syntactic valency, and distributional regularity. It is well suited as a bounded case. Its analytical steps are enumerable: core metrics such as mean dependency distance, type-token ratio, and Zipf exponent form a mature and finite repertoire that maps cleanly onto a library of reusable processing nodes. Each node carries an explicit computational definition. Metric nodes come with their mathematical formula, and other nodes come with algorithmic descriptions and input-output specifications. The input and output of each step are also formally typed. Parsed treebanks go in, numerical summaries or statistical models come out, and nodes can therefore enforce deterministic data-flow contracts at the interface. In addition, the field has a measurement tradition spanning several decades \cite{liu2008,lu2010,zipf1949}, which supplies both the analytical vocabulary and the reference values needed for automated output validation. Beyond this technical suitability, quantitative syntax has a concrete practical need: the community needs a standardised, shareable resource package, including a domain-specific node library, benchmark task definitions, and workflow templates. With such a package in place, researchers can spend their effort on theoretical interpretation instead of getting stuck in implementation details.

Given these observations, we address three research questions:

\textbf{RQ1.} Can natural-language research descriptions be transformed, via AI-assisted construction, into visual and executable workflow representations that make the underlying analytical logic inspectable and reproducible?

\textbf{RQ2.} Can shared workflow objects be incrementally refined by researchers to accommodate evolving analytical requirements, thereby extending their lifecycle as reusable language resources?

\textbf{RQ3.} What methodological costs, boundaries, and applicability conditions accompany this workflow-centred approach to language research?

RQ1 tests whether the \textit{representation gap} identified above can be bridged through a structured generation pipeline that converts free-text task descriptions into deterministic, executable workflows. RQ2 examines whether the resulting workflow objects, once saved and shared, can serve as living research artefacts that evolve with their users' needs rather than being discarded and rebuilt from scratch. RQ3 situates the approach within its practical limits, asking what trade-offs in computational cost, benchmark scope, and domain specificity must be acknowledged.

The paper makes three contributions. The first is methodological evidence: AI-assisted workflow construction and incremental refinement are viable within a bounded research domain. On a 64-task benchmark grounded in quantitative-syntax research traditions, the proposed pipeline achieves full structural and executability compliance, with a mean output-plausibility rate of 98.4\% across three independent runs. On a complementary 12-task lifecycle benchmark (36 trials across three runs), patch-based refinement succeeds in every case and uses roughly one-third of the tokens needed by full regeneration. The second is a set of reusable language-research resources: a domain-specific node library (48 nodes across nine categories of quantitative-syntax operations), a literature-driven benchmark suite called QL-Bench (10 research directions, 3 difficulty levels, 64 tasks in total), and exportable workflow templates in a standard JSON format. All three resources are integrated into QLWF, a workflow platform for quantitative linguistics, and can be accessed, composed, and extended. The third is a workflow-based evaluation framework. Within this framework, the three-level assessment protocol (structural validity, executability, and output plausibility) provides a controlled instrument for comparing alternative workflow-generation strategies. AI assistance is confined to the construction stage. Judgement on the workflow's adequacy stays with the researcher.

We organise the remainder of this paper as follows. Section~\ref{sec:related-work} reviews related work and identifies the gap addressed. Section~\ref{sec:method} presents the workflow-centred methodology. Section~\ref{sec:experiments} reports experiments and results. Section~\ref{sec:discussion} discusses implications and limitations. Section~\ref{sec:conclusion} concludes.


\section{Related Work}\label{sec:related-work}

Several platforms have been developed for constructing and executing language-processing workflows. TextFlows \cite{perovsek2016} is a web-based platform that enables visual composition, execution, and sharing of text mining workflows. It is the closest existing model to the approach explored in this paper, as it combines a drag-and-drop workflow editor with public sharing via unique URLs. WebLicht \cite{hinrichs2010}, developed within the CLARIN infrastructure, provides web-based chaining of NLP services for corpus annotation, though its workflows cannot be freely published for community reuse. Recent LRE work on Research-Infrastructure-as-a-Service likewise emphasises web-based access to language-processing services and task-oriented chaining, but not AI-assisted workflow construction \cite{gomes2025}. GATE \cite{cunningham2002} offers a mature suite of NLP tools with a graphical interface, but its pipeline-sharing capabilities are limited to institutional platforms such as GATE Cloud. Other platforms such as LAPPS Grid \cite{ide2014} and OpenMinTeD \cite{labropoulou2018} provide web-service architectures for connecting NLP components, but neither offers a visual workflow editor oriented towards linguistic research. KNIME \cite{berthold2009} is the most widely adopted general-purpose workflow platform and includes basic NLP components via plugins, but provides no domain-specific node library tailored to linguistic analysis. Table~\ref{tab:platform-comparison} summarises these platforms. A consistent gap emerges: existing platforms support manual workflow construction to varying degrees, but none provides AI-assisted construction from natural-language research descriptions.

\begin{table}[htbp]
\caption{Comparison of language-processing workflow platforms along dimensions relevant to the three research questions.}\label{tab:platform-comparison}
{\footnotesize
\begin{tabular*}{\textwidth}{@{\extracolsep\fill}lllll}
\toprule
Platform & Construction & Revision & Domain library & Sharable \\
\midrule
TextFlows\textsuperscript{a}    & Manual       & Manual re-editing & NLP widgets  & Public URL \\
WebLicht\textsuperscript{b}     & Manual       & Not supported     & NLP services & Limited \\
GATE\textsuperscript{c}         & Manual       & Manual re-editing & NLP plugins  & Limited \\
LAPPS Grid\textsuperscript{d}   & Manual       & Not supported     & NLP services & Yes \\
KNIME\textsuperscript{e}        & Manual       & Manual re-editing & Via plugins  & Yes \\
InstructPipe\textsuperscript{f} & AI-assisted  & Not supported     & None         & No \\
QLWF (this paper)               & AI-assisted  & Incremental patch & 48 nodes     & JSON \\
\bottomrule
\end{tabular*}
\footnotetext{Note: ``Manual'' denotes hand assembly in a visual editor. ``AI-assisted'' denotes workflows generated from a natural-language description. ``Incremental patch'' denotes targeted revision of a saved workflow object.}
\footnotetext{\textsuperscript{a}\cite{perovsek2016}; \textsuperscript{b}\cite{hinrichs2010}; \textsuperscript{c}\cite{cunningham2002}; \textsuperscript{d}\cite{ide2014}; \textsuperscript{e}\cite{berthold2009}; \textsuperscript{f}\cite{zhou2025}.}
}
\end{table}

Other work has explored AI-assisted generation of workflows from natural-language descriptions. InstructPipe \cite{zhou2025} is the most directly comparable system: it uses a large language model to convert user instructions into visual pipelines within the Visual Blocks framework. However, InstructPipe targets general-purpose machine learning tasks and provides neither a domain-specific node library nor a mechanism for revising previously generated pipelines. Agent-workflow systems such as AFlow \cite{zhang2025} and Meta-Agent-Workflow \cite{tan2025} take a different direction. These systems construct workflows whose execution paths involve runtime decisions by a language model, offering adaptability but sacrificing the determinism required for reproducible empirical research. At a broader level, natural-language-to-code systems can produce analysis scripts or notebook code from textual prompts \cite{yin2023}. But the resulting code is not structured as an inspectable workflow object and therefore does not address the representation gap described in Section~\ref{sec:introduction}. The limitation is consistent across these approaches: existing AI-assisted systems either target general domains without linguistic depth, or produce agent workflows that lack deterministic execution guarantees. None combines AI-assisted construction with domain-specific depth and incremental refinement of the generated artefacts.

Beyond construction, the broader scientific-workflow community has increasingly focused on how workflow artefacts are maintained and revised over time. The Common Workflow Language (CWL) \cite{crusoe2022} established a platform-independent standard for describing computational workflows, and the FAIR Workflows initiative \cite{goble2020} extended the FAIR principles (Findable, Accessible, Interoperable, Reusable) to workflow objects themselves. WorkflowHub \cite{gustafsson2025} provides a registry through which researchers can discover, cite, and version scientific workflows. At the level of community practice, nf-core \cite{ewels2020} demonstrates how a curated collection of bioinformatics pipelines can be maintained through collaborative version control. On the analytical side, Missier \cite{missier2016} proposed PDIFF, a provenance-differencing algorithm that tracks structural changes between successive versions of a workflow. These efforts show that workflow revision is a recognised need in data-intensive sciences. However, in language research, incremental refinement of shared workflow objects has not yet been systematically addressed. Our work draws on the insight from this literature that workflow artefacts, like other research outputs, benefit from structured revision mechanisms, and applies it within the bounded domain of quantitative syntax.

A final question is whether workflows have been explicitly framed as \textit{language resources} or as \textit{instruments for evaluating research procedures}. TextFlows comes closest: its public workflows function as shareable research artefacts, and the platform has been used to compare NLP components under controlled conditions \cite{martinc2024}. Related LRE work on reproducibility and reuse has also shown that specific NLP procedures can be packaged as shareable visual workflows in ClowdFlows, making them easier to reuse outside the original experimental setting \cite{repar2020}. Yet these studies do not explicitly theorise the workflow as a resource object in the sense of LRAE, nor do they propose a structured evaluation protocol for comparing workflow-generation strategies. In the broader workflow community, WorkflowHub \cite{gustafsson2025} treats workflows as FAIR digital objects with persistent identifiers, but it serves as a registry rather than an execution or evaluation environment. To our knowledge, no existing work has combined the resource perspective with an evaluation framework in which workflow objects themselves serve as the medium for comparing alternative generation or analysis strategies.

The literature reviewed above reveals progress along each of the three dimensions addressed by our research questions, but no single body of work spans all three. Existing language-processing platforms support manual workflow construction and, in some cases, sharing, yet none offers AI-assisted generation from natural-language research descriptions. AI-assisted workflow generation has advanced rapidly, but current systems either target general domains without linguistic depth or produce agent workflows that lack deterministic execution. Scientific-workflow research has established principles and tools for workflow versioning and revision, but these have not been applied to language-research workflows. Finally, while individual platforms have touched on the resource and evaluation potential of workflows, no prior work has proposed a unified framework in which workflow objects serve simultaneously as reusable language resources and as instruments for evaluating research procedures. We address this combined gap within the bounded domain of quantitative syntax.


\section{A Workflow-Centred Approach to Language Research}\label{sec:method}

\subsection{Workflow as a Language-Research Object}\label{sec:workflow-object}

\begin{figure}[htbp]
\centering
\includegraphics[width=\textwidth]{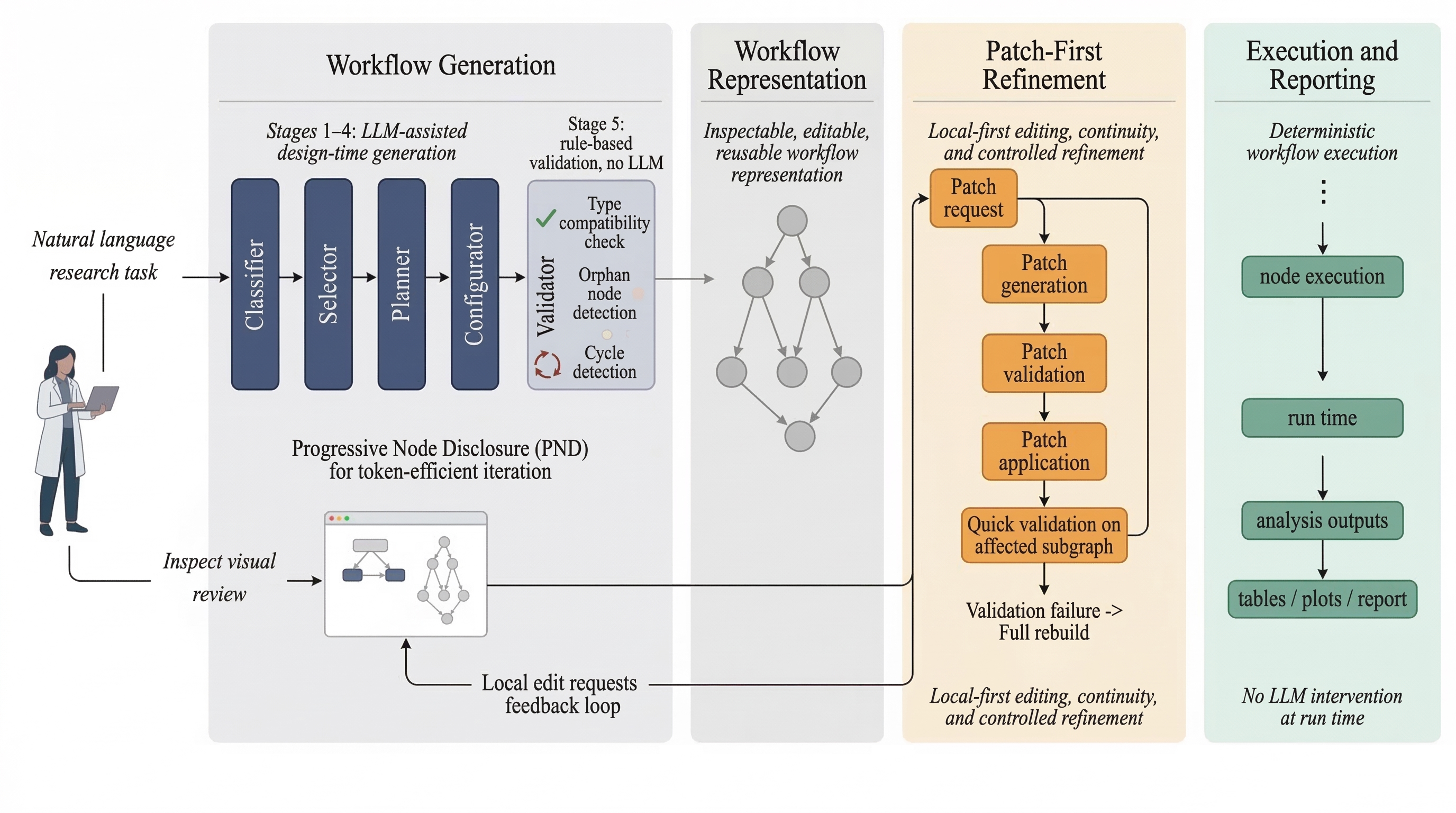}
\caption{System overview of the workflow-centred approach. Research intent is first reified as a visual node-flow structure and then formalised into a deterministic, executable data pipeline.}\label{fig:system-overview}
\end{figure}

The methodological framework of this paper rests on a dual transformation that we term \textit{reification and formalization}. Figure~\ref{fig:system-overview} illustrates the overall approach. We borrow the concept of reification from Wenger \cite{wenger1998}, who defines it as ``the process of giving form to our experience by producing objects that congeal this experience into `thingness'\,'' (p.~58). In our context, reification refers to the process by which a researcher's implicit analytical logic is externalised as a visible workflow structure: a graph of named processing nodes connected by explicit data-flow edges. This transformation makes the research logic available for inspection, discussion, and sharing. However, reification alone does not guarantee that the resulting structure can be executed. A second layer, \textit{formalization}, is therefore needed, by which we mean the transformation of a visible but informal workflow sketch into a structure with precise input-output type contracts and deterministic execution semantics. In computational linguistics, formalization in this sense has a long tradition (cf.\ \cite{montague1970}). The combination of these two layers means that a single workflow object addresses two needs: it is human-readable as a visual node graph that a researcher can inspect and modify, and it is machine-executable as a deterministic data pipeline that a computer can run with reproducible results.

We define the product of formalization as a workflow object
\begin{equation}
\label{eq:workflow-object}
W = (N, E, P),
\end{equation}
where $N$ is a set of processing nodes, $E$ is a set of directed edges representing data-flow dependencies, and $P$ is a parameter mapping that associates each node with its configuration values. This definition gives the workflow a precise mathematical identity. Table~\ref{tab:symbols} summarises the core symbol conventions used throughout this paper. The workflow is not a flowchart sketch or a loose diagram, but a structured object whose components can be enumerated, compared, and serialised. Because every node declares typed input and output ports, the graph structure enforces compatibility constraints at design time. A workflow that passes these constraints is guaranteed to be executable without runtime type errors. The formal definition thus serves as the foundation for both automated validation and deterministic execution.

\begin{table}[htbp]
\caption{Symbol conventions used throughout the paper.}\label{tab:symbols}
\begin{tabular*}{\textwidth}{@{\extracolsep\fill}ll}
\toprule
Symbol & Meaning \\
\midrule
$W = (N, E, P)$ & Dataflow workflow: node set $N$, edge set $E$, parameter map $P$ \\
$G_\pi(q)$ & QLWF five-stage pipeline generation result for query $q$ \\
$G_m(t)$ & Workflow generated by method $m$ for task $t$ \\
$L_k(W, t)$ & Level-$k$ evaluation metric ($k \in \{1, 2, 3\}$) \\
$\text{Pass@1}_k$ & First-attempt success rate on the task set \\
$\mathcal{T}$ & Benchmark task set used for level-wise aggregation \\
\bottomrule
\end{tabular*}
\end{table}

Reification becomes precise only when the available building blocks correspond to recognisable research operations. The QLWF node library is organised into 48 nodes across nine categories: corpus preprocessing, lexical analysis, syntactic analysis, statistical modelling, L2 syntactic complexity (following \cite{lu2010}), visualisation, type conversion, data input, and data output. Table~\ref{tab:node-ecosystem} provides an overview of each category, its scope, and representative node names. The library covers the core metrics of quantitative syntax, including mean dependency distance, type-token ratio, Zipf-rank analysis, and 14 syntactic complexity indices from the L2SCA tradition. Each node encapsulates a single, well-defined research operation with typed input and output ports. Researchers compose workflows by connecting these nodes on a visual canvas, thereby externalising their analytical logic as an explicit graph structure. In practice, the node library serves a dual role: it is both the vocabulary through which reification is expressed and a lightweight domain ontology that encodes the operational categories of quantitative-syntax research.

\begin{table}[htbp]
\caption{QLWF node library: 48 nodes across nine categories.}\label{tab:node-ecosystem}
\begin{tabular*}{\textwidth}{@{\extracolsep\fill}lr p{5.5cm} p{3cm}}
\toprule
Category & Nodes & Scope & Representative nodes \\
\midrule
Corpus & 5 & Tokenisation, text cleaning, stopword filtering, word frequency, N-gram extraction & Tokenizer, WordFrequency \\
Lexical & 5 & Lexical diversity and richness (five indices) & Type-token ratio, moving-average TTR, lexical diversity metric \\
Syntactic & 6 & Dependency distance, hierarchy, tree metrics, valency, direction & MDD, MHD, Valency \\
Statistical & 4 & Entropy, Zipf-rank analysis, correlation, descriptive statistics & Entropy, ZipfAnalysis \\
L2SCA & 2 & 14 syntactic complexity indices & L2SCAMetrics \\
Visualisation & 7 & Frequency charts, Zipf curves, histograms, word clouds, scatter plots & FrequencyChart, WordCloud \\
Converter & 11 & Bidirectional type adaptation across seven port types & UniversalConverter \\
Input & 4 & Interactive and file-based data ingestion & TextInput, FileInput \\
Output & 4 & Tabular display, JSON preview, file export & TableOutput, FileExport \\
\midrule
Total & 48 & & \\
\bottomrule
\end{tabular*}
\end{table}

\begin{figure}[htbp]
\centering
\includegraphics[width=\textwidth]{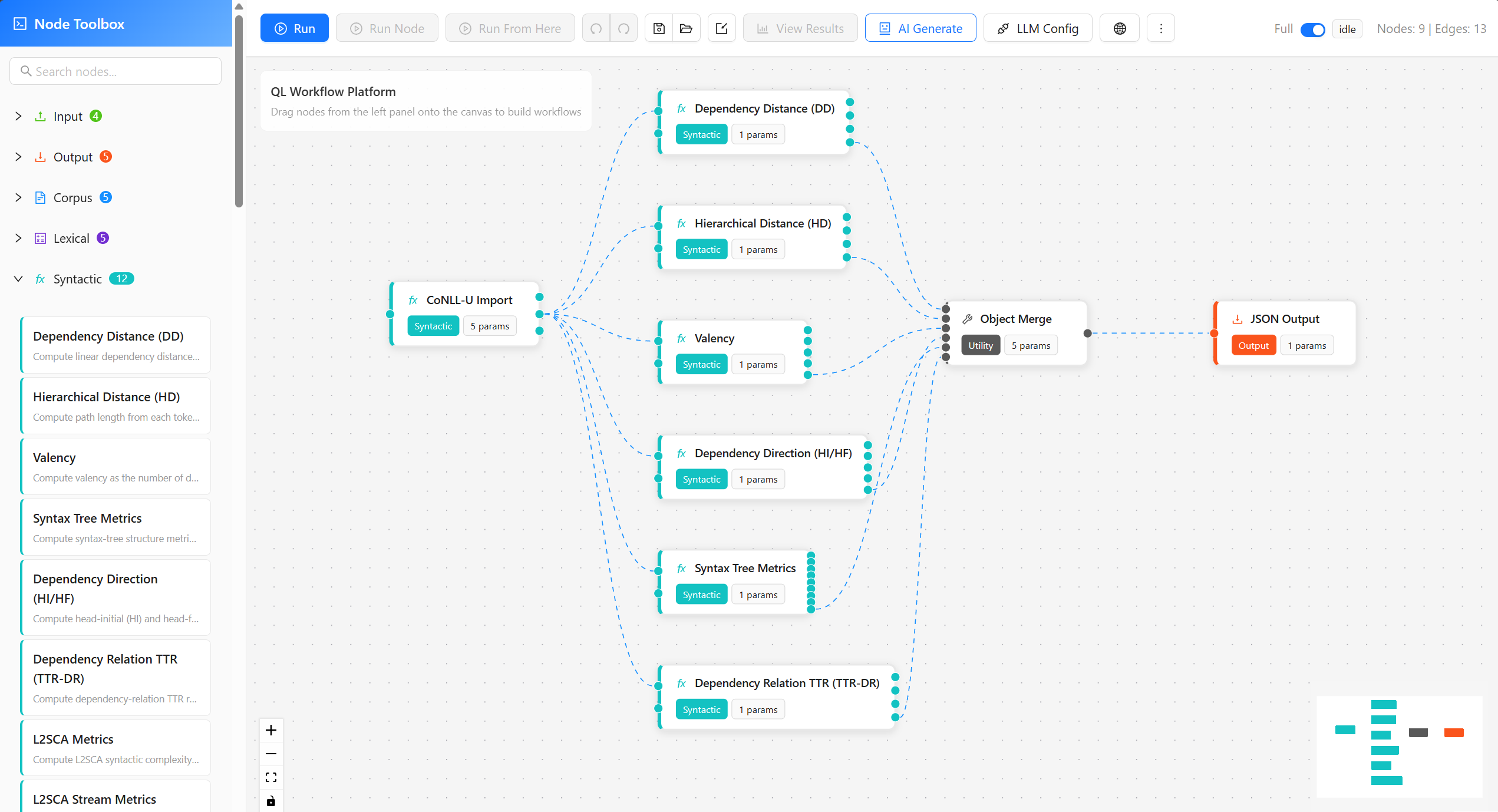}
\caption{The QLWF editing environment. The workflow graph is displayed as an interactive canvas on which researchers can construct, inspect, and modify analytical pipelines.}\label{fig:editing-environment}
\end{figure}

Once constructed, a workflow object can be persisted, exported, and shared as a reusable language-research artefact. We store workflows in a standard JSON format that captures the full graph structure: node types, parameter settings, edge connections, and layout coordinates. Researchers can export a workflow as a self-contained JSON file and share it with collaborators, who can import it into their own workspace, inspect its structure, modify its parameters, and re-execute it. Figure~\ref{fig:editing-environment} shows the QLWF editing environment, in which the workflow graph is displayed as an interactive canvas. This persistence and exchange mechanism is analogous to the public workflow URLs offered by TextFlows \cite{perovsek2016} and aligns with the FAIR Workflows principle that computational workflows should be findable, accessible, and reusable as first-class research objects \cite{goble2020}. The resource property of workflows is designed in: the same object that a researcher constructs for analysis can be handed to another for replication, adaptation, or extension.

\FloatBarrier
\subsection{From Research Intent to Executable Workflow}\label{sec:construction-pipeline}

One design boundary is central to the construction pipeline. In QLWF, the large language model participates only at \textit{design time}: it assists the researcher in assembling the workflow structure. Once the workflow enters execution, all computation is carried out by hard-coded node implementations with fixed input-output contracts. No language model is involved at runtime. The stability of this deterministic data flow is what ensures that results are reproducible. This boundary distinguishes our approach from agent-workflow systems in which a language model makes runtime decisions during execution \cite{zhang2025,tan2025}.

With this boundary in place, the five-stage construction pipeline, denoted
\begin{equation}
\label{eq:construction-pipeline}
G_\pi(q) = \left(f_\text{val} \circ f_\text{cfg} \circ f_\text{plan} \circ f_\text{sel} \circ f_\text{cls}\right)(q),
\end{equation}
transforms a natural-language task description $q$ into an executable workflow. (1)~\textit{Classify} identifies the research intent and selects a task category. (2)~\textit{Select} retrieves candidate nodes from the domain-specific library. (3)~\textit{Plan} determines the execution order and data-flow connections. (4)~\textit{Configure} assigns parameter values to each node. (5)~\textit{Validate} checks structural and type-level consistency before the workflow is released for execution. Each stage receives a structured context object from the previous stage and produces a refined context for the next. In terms of the reification-formalization framework, the pipeline realises a progressive transformation: research intent is gradually externalised into concrete node selections (reification) while constraints are progressively narrowed into a deterministic structure (formalization). The researcher retains authority over the final product: the generated workflow can be inspected, edited, or rejected before execution.

Each stage produces a typed output that constrains the input space of the next stage. For example, the classify stage outputs a task category and a set of relevant research directions, which the select stage uses to narrow the candidate node pool. The plan stage then receives only the selected nodes and must connect them into a valid data-flow graph. Determinism is therefore a property not just of the final workflow but of the entire construction process. At no point does any stage receive unconstrained free text from the previous stage.

\subsection{Incremental Refinement of Workflow Objects}\label{sec:incremental-refinement}

Saved workflow objects are not disposable outputs. They are research artefacts that embody verified analytical logic. Discarding them means losing that verification. When research requirements change, full reconstruction from scratch is wasteful because it discards the validated structure of the existing workflow. We address this problem through incremental refinement, which applies targeted modifications to a saved workflow while preserving its overall integrity. The same concern appears in the broader scientific-workflow literature on workflow revision \cite{missier2016,ewels2020}, though the application here is specifically to language-research workflows.

\begin{figure}[htbp]
\centering
\includegraphics[width=\textwidth]{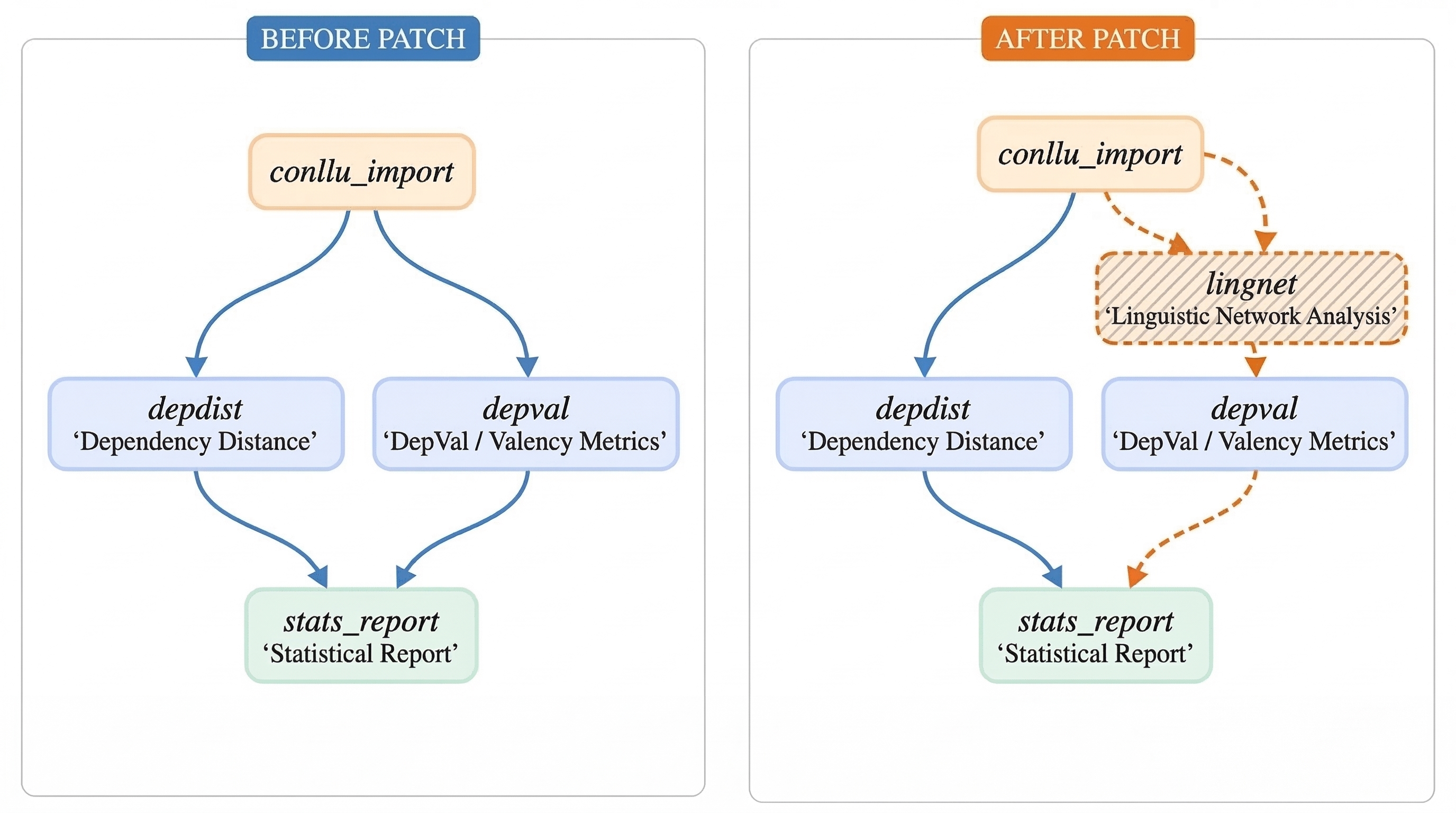}
\caption{Example of patch-based refinement. Left: original workflow. Right: modified workflow after a targeted patch operation. Only the affected nodes and edges are changed. The rest of the verified structure is preserved.}\label{fig:patch-example}
\end{figure}

We support four types of refinement operations, each corresponding to a common research scenario. A \textit{parameter change} adjusts a configuration value within an existing node, for example changing the corpus selection or a statistical threshold. A \textit{node insertion} adds a new analytical step, such as appending a valency analysis node to an existing dependency-distance workflow. A \textit{composite edit} reorganises part of the analysis chain, for example replacing one preprocessing strategy with another. A \textit{rewiring} modifies the data-flow connections between existing nodes, such as redirecting output from a single visualisation to a multi-corpus comparison. Figure~\ref{fig:patch-example} shows an example of a workflow before and after a patch-based refinement operation.

After a refinement operation is applied, the modified workflow undergoes the same three-level validation used for newly constructed workflows: structural validity ($L_1$), executability ($L_2$), and output plausibility ($L_3$). This ensures that the refinement has not introduced inconsistencies. The validation framework checks both the correctness of the reified structure (are the nodes and edges well-formed?) and the correctness of the formalised execution (does the workflow run and produce plausible output?). This approach preserves the verified parts of the original workflow, modifying only what needs to change. The resulting workflow objects, together with the node library and benchmark tasks, constitute a reusable set of language-research artefacts.

\subsection{A Typical Usage Scenario}\label{sec:usage-scenario}

A typical QLWF session arises in day-to-day research practice rather than on a benchmark.

Consider a researcher in quantitative syntax who wants to compare the syntactic profile of English and Chinese on several indicators at once. The researcher opens QLWF and drags a CoNLL-U Import node from the Node Toolbox onto the canvas. Six syntactic nodes follow: Dependency Distance, Hierarchical Distance, Valency, Dependency Direction, Syntax Tree Metrics, and Dependency Relation TTR. To collect their outputs, the researcher connects them into an Object Merge node, which feeds a JSON Output. The resulting pipeline is the one shown in Figure~\ref{fig:editing-environment}. A single click on Run executes the workflow. The View Results panel returns the merged indicator vector. At this point the workflow can be saved and shared in the JSON form described in Section~\ref{sec:workflow-object}, which means a second researcher can reproduce the session without rebuilding anything.

A second path is open to the same researcher. Instead of building the pipeline node by node, the researcher can click AI Generate in the top toolbar and type a natural-language description, for instance ``compare syntactic complexity across English-EWT and Chinese-GSD on six indicators''. The five-stage pipeline from Section~\ref{sec:construction-pipeline} returns a candidate workflow on the canvas, all at design time only. From there the path is the same as before. The researcher can accept the draft, adjust parameters, add or remove nodes, and then run it. Whether the workflow is hand-built or AI-generated, the execution step itself does not change.

One researcher, working through one session, walks away with something more durable than a one-off analysis: a workflow object that can serve as a starting point for later studies.


\section{Experiments and Results}\label{sec:experiments}

The evaluation tests whether workflow-as-object is a viable methodological instrument, not whether the underlying system is performant. Section~\ref{sec:setup} describes the experimental setup, followed by results on workflow construction feasibility (Section~\ref{sec:construction-results}), incremental refinement (Section~\ref{sec:refinement-results}), and methodological cost (Section~\ref{sec:cost}).

\subsection{Experimental Setup}\label{sec:setup}

\begin{figure}[htbp]
\centering
\includegraphics[width=\textwidth]{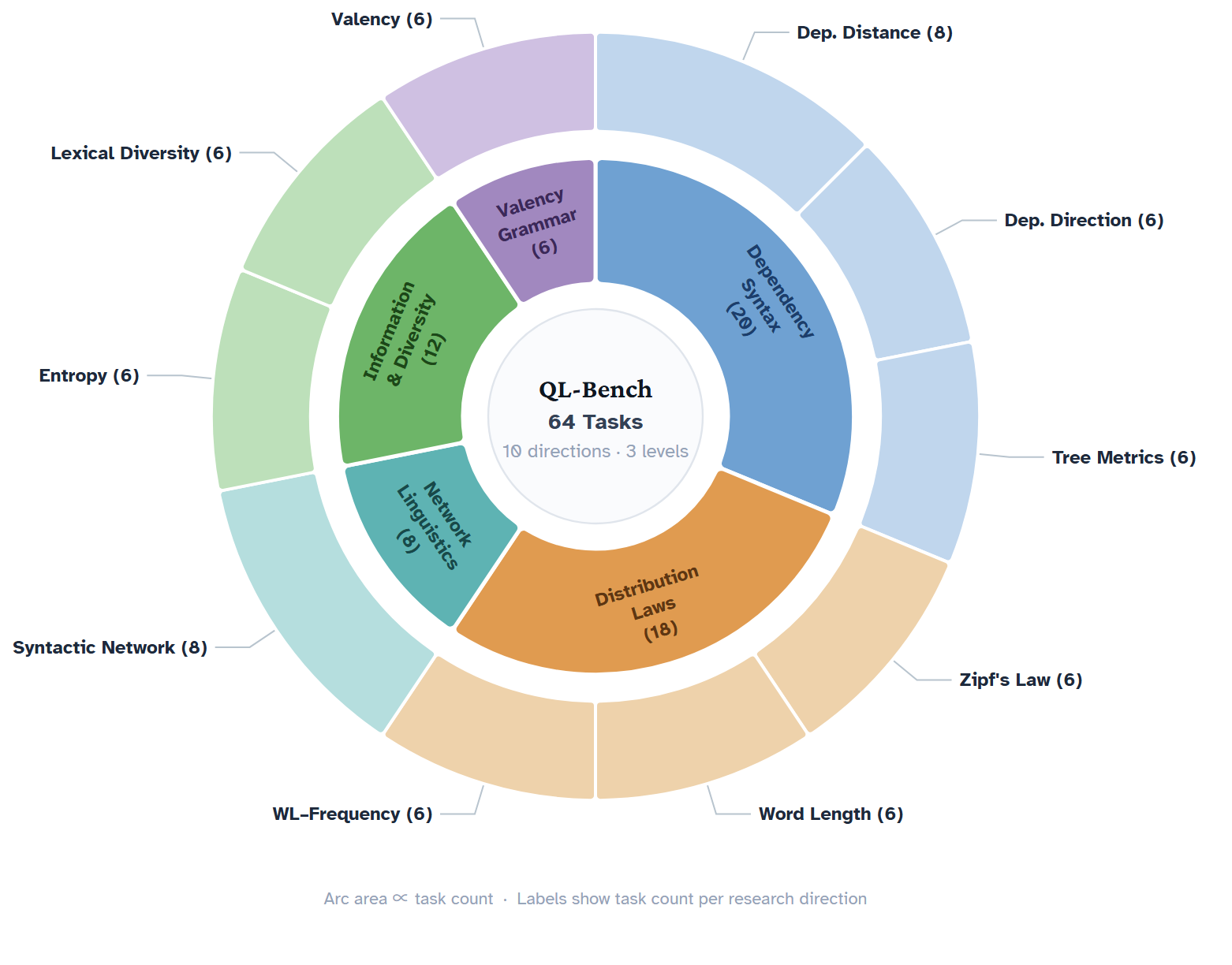}
\caption{Provenance structure of QL-Bench. The 64 tasks are derived from five quantitative-syntax research traditions, organised into ten research directions at three difficulty levels.}\label{fig:provenance}
\end{figure}

The benchmark used in this study, QL-Bench, is grounded in the quantitative-syntax literature. Its tasks are derived from five established research traditions: dependency syntax, distributional laws, syntactic networks, information and diversity measures, and valency grammar. From these traditions we identified a set of core research directions and designed tasks at three difficulty levels. Easy tasks involve a single metric applied to a single corpus. Medium tasks require multiple metrics or cross-corpus comparison. Hard tasks involve multi-step analysis chains or statistical model fitting. QL-Bench is a researcher-constructed local benchmark, not a community-endorsed standard. Its purpose is to provide a controlled, literature-grounded testbed for evaluating workflow construction. Figure~\ref{fig:provenance} illustrates the provenance structure.

QL-Bench comprises 64 tasks spanning the ten research directions and three difficulty levels. The tasks collectively involve 24 node types from the QLWF library and draw on three Universal Dependencies treebanks: Chinese-GSD, English-EWT, and Chinese-PUD \cite{nivre2020}. Some tasks reference more than one corpus, resulting in 85 total corpus citations across the 64 tasks. Each task is defined by a natural-language input description, an expected node combination, and a set of $L_3$ validation rules.

We evaluate workflow quality using a three-level protocol. For a task $t$ and generated workflow $W$, the three levels can be expressed as
\begin{equation}
\label{eq:l123-definition}
L_k(W,t)=
\begin{cases}
\mathds{1}\!\left[\operatorname{expected}(t)\subseteq \operatorname{types}(W)\ \land\ \operatorname{wellformed}(W)\right], & k=1,\\[4pt]
\mathds{1}\!\left[\operatorname{run}(W)=\texttt{success}\right], & k=2,\\[4pt]
\mathds{1}\!\left[\forall c \in \operatorname{checks}(t): c\!\left(\operatorname{out}(W)\right)\right], & k=3.
\end{cases}
\end{equation}
Here, $\operatorname{expected}(t)$ denotes the required node types for task $t$, $\operatorname{types}(W)$ the node types realized in $W$, and $\operatorname{checks}(t)$ the task-specific output constraints. In interpretive terms, $L_1$ captures structural validity, $L_2$ executability, and $L_3$ output plausibility. In terms of the reification-formalization framework introduced in Section~\ref{sec:workflow-object}, $L_1$ tests whether research intent has been correctly externalised as a node-flow structure, $L_2$ tests whether that structure has valid execution semantics, and $L_3$ tests end-to-end effectiveness.

The $L_3$ assessment is fully automated and free from subjective judgement. Each benchmark task specifies a set of validation rules at definition time. These rules include three types of check. (1)~Designated output fields must be non-empty (e.g., the dependency distance list). (2)~Numerical values must fall within literature-derived ranges (e.g., mean dependency distance between 2.0 and 5.0). (3)~Certain values must be positive (e.g., entropy). The 64 tasks collectively employ nine distinct check types. All checks must pass for a task to receive an $L_3$ pass. No human judgement is involved at any point. Because we fixed the acceptance thresholds before the experiment began and derived them from published quantitative-syntax results, the evaluation is free from post-hoc adjustment. This design ensures that the $L_3$ assessment measures output plausibility against an objective, pre-registered standard rather than relying on case-by-case expert evaluation. For a generation method $m$ evaluated on benchmark task set $\mathcal{T}$, the level-wise first-attempt score is then defined as
\begin{equation}
\label{eq:pass-at-one}
\mathrm{Pass@1}_k = \frac{1}{|\mathcal{T}|}\sum_{t \in \mathcal{T}} L_k\!\left(G_m(t), t\right),
\qquad k \in \{1,2,3\}.
\end{equation}

We compare four generation strategies, all of which take the same natural-language task description as input. The three baselines differ in how the language model processes this input. \textit{Single-prompt} passes the description directly to the language model and requests a complete workflow in a single response. \textit{Chain-of-thought} adds a step-by-step reasoning preamble to the same single-call setup, so that the model verbalises its analytical plan before producing the workflow. \textit{Two-stage} separates planning from generation into two independent calls: the first produces a node-level plan, and the second takes that plan as input and generates the workflow structure. The fourth strategy is the QLWF pipeline described in Section~\ref{sec:construction-pipeline}, which processes the same input through five structured stages with domain-specific context at each stage. These four strategies represent a progression from minimal structure to fully constrained, multi-stage construction, allowing us to assess the contribution of pipeline structure to workflow quality.

To evaluate incremental refinement (RQ2), we use a separate 12-task lifecycle benchmark. Each task consists of an initial workflow and a modification request that changes the analytical requirements. The modification types correspond to the four refinement operations defined in Section~\ref{sec:incremental-refinement}. This benchmark tests whether a saved workflow can be successfully revised through targeted patches rather than rebuilt from scratch.

All experiments use GLM-5, a model from the GLM family \cite{du2022}, accessed via the Zhipu AI API. The three UD treebanks are held fixed across all conditions. Each task in the 64-task benchmark is run under all four generation strategies using identical prompts and configurations. To assess stability, each condition is repeated three times. The reported figures are averaged across runs. Token counts are approximate values recorded by the evaluation runner and should not be interpreted as vendor-level billing figures. All four generation strategies were evaluated in parallel against the same API endpoint. Latency figures are therefore approximate and should not be used for precise cross-method speed comparisons.

\subsection{Workflow Construction Feasibility}\label{sec:construction-results}

\begin{figure}[htbp]
\centering
\includegraphics[width=\textwidth]{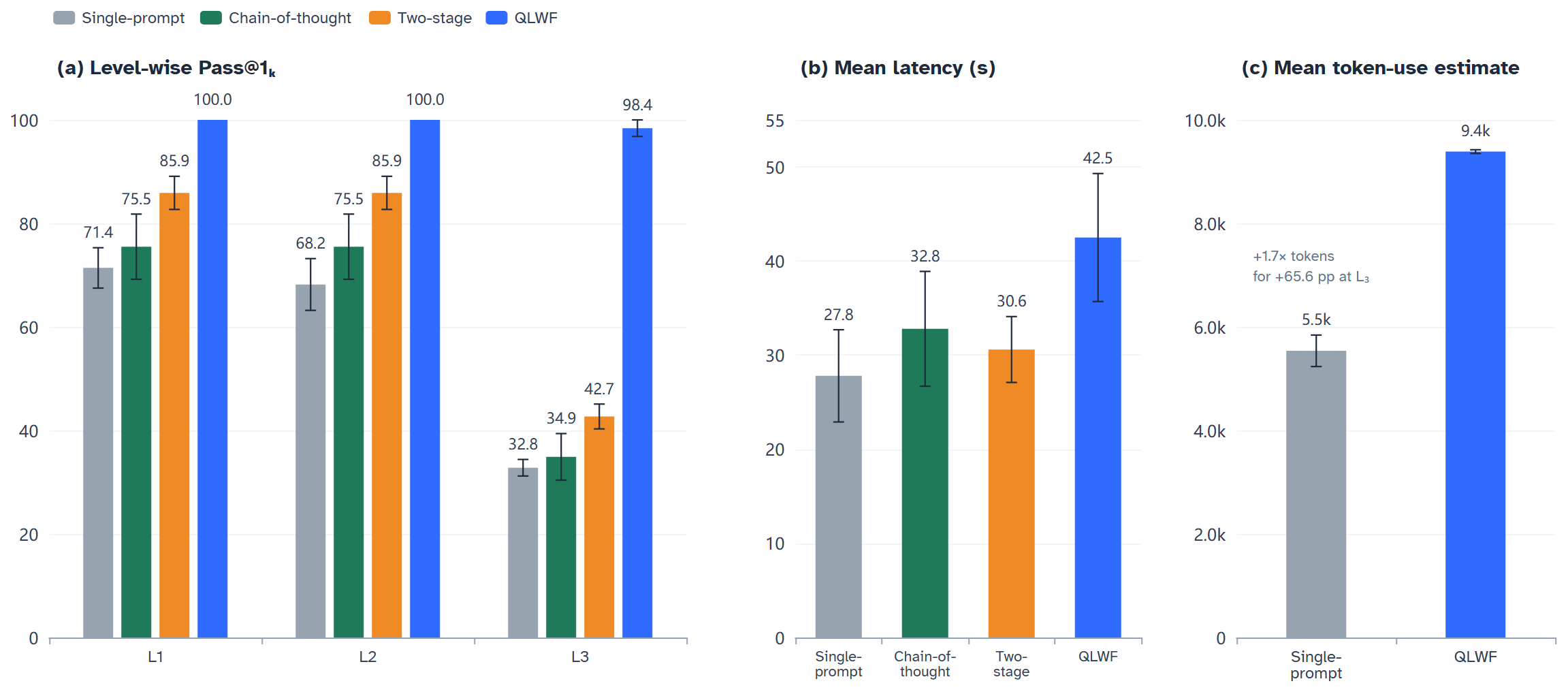}
\caption{QL-Bench main results across four generation strategies on 64 tasks. (a) Pass@1$_k$ at evaluation levels $L_1$ (structural validity), $L_2$ (executable), and $L_3$ (output validity). (b) Mean inference latency in seconds (approximate, parallel API execution). (c) Mean token-use estimate per task. Only Single-prompt and QLWF return usable token counts. Error bars show $\pm$1 SD across 3 independent runs (values tabulated in Table~\ref{tab:main-results}).}\label{fig:main-results}
\end{figure}

Table~\ref{tab:main-results} summarises the main results. QLWF achieves $L_1$ = 100.0\%, $L_2$ = 100.0\%, and $L_3$ = 98.4\% (mean across three runs, from individual runs of 62, 63, and 64 out of 64). The three prompt-based baselines show considerably lower pass rates, particularly at the $L_3$ level: Single-prompt reaches 32.8\%, Chain-of-thought 34.9\%, and Two-stage 42.7\%. Figure~\ref{fig:main-results}(a) visualises these results. Within the scope of the current benchmark, these results suggest that AI-assisted workflow construction through a structured pipeline is feasible.

\begin{table}[htbp]
\caption{QL-Bench main results across four generation methods and three evaluation levels ($n = 64$ tasks). Values are mean $\pm$ std across 3 independent runs. Latency is approximate (parallel execution). Token counts marked ``--'' were not returned by the API for these methods.}\label{tab:main-results}
\begin{tabular*}{\textwidth}{@{\extracolsep\fill}lrr@{\,$\pm$\,}lr@{\,$\pm$\,}lr@{\,$\pm$\,}lr@{\,$\pm$\,}lr@{\,$\pm$\,}l}
\toprule
Method & Tasks & \multicolumn{2}{c}{$L_1$ (\%)} & \multicolumn{2}{c}{$L_2$ (\%)} & \multicolumn{2}{c}{$L_3$ (\%)} & \multicolumn{2}{c}{Avg.\ tokens} & \multicolumn{2}{c}{Lat.\ (s)} \\
\midrule
Single-prompt    & 64 & 71.4  & 3.9 & 68.2  & 5.0 & 32.8 & 1.6 & 5{,}539 & 304 & 27.8 & 4.9 \\
Chain-of-thought & 64 & 75.5  & 6.3 & 75.5  & 6.3 & 34.9 & 4.5 & \multicolumn{2}{c}{--} & 32.8 & 6.1 \\
Two-stage        & 64 & 85.9  & 3.2 & 85.9  & 3.2 & 42.7 & 2.4 & \multicolumn{2}{c}{--} & 30.6 & 3.5 \\
QLWF             & 64 & 100.0 & 0.0 & 100.0 & 0.0 & 98.4 & 1.6 & 9{,}387 & 36  & 42.5 & 6.8 \\
\bottomrule
\end{tabular*}
\end{table}

At $L_1$, the gap between QLWF and the strongest baseline (Two-stage, 85.9\%) is already substantial but could be attributed in part to engineering advantages. At $L_3$, however, the gap widens dramatically: 98.4\% versus 42.7\%. This gap shows that the pipeline's advantage extends beyond assembling correct node structures. It produces workflows whose outputs are actually valid. Prompt-level reasoning can improve structural quality, but it is not, on its own, sufficient to guarantee output correctness.

This stronger output validity comes at a limited additional methodological cost. QLWF uses an average of 9,387 tokens per task compared with 5,539 for Single-prompt, a ratio of approximately 1.7. Average latency is 42.5 seconds for QLWF versus 27.8 seconds for Single-prompt. Figures~\ref{fig:main-results}(b) and \ref{fig:main-results}(c) show the latency and token-use distributions with $\pm$1 SD error bars. As noted above, latency was measured under parallel execution and is reported as an approximate indicator rather than a precise comparison. The multi-stage pipeline architecture queries the language model at each of its five stages, which explains the higher token count. QLWF also shows higher latency under the parallel evaluation setup, though this difference may partly reflect API contention rather than intrinsic processing time.

\begin{figure}[htbp]
\centering
\includegraphics[width=\textwidth]{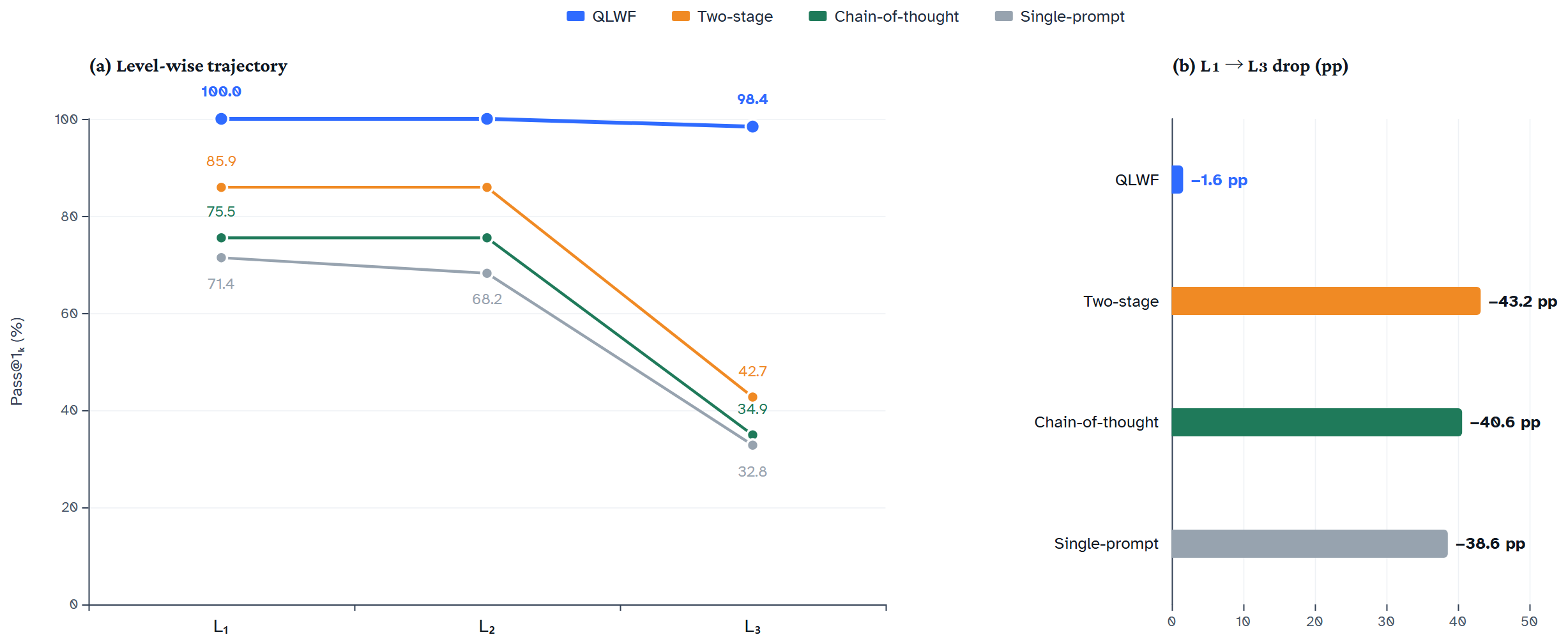}
\caption{Cascade view of QL-Bench results. (a) Level-wise Pass@1$_k$ (\%) trajectory across $L_1$, $L_2$, and $L_3$ for each generation strategy (3-run means on 64 tasks, with standard deviations in Table~\ref{tab:main-results}). (b) Cumulative $L_1 \to L_3$ drop per method. QLWF loses only $1.6$ pp between structural validity and output validity, whereas the three prompt-based baselines lose between $38.6$ and $43.2$ pp.}\label{fig:cascade}
\end{figure}

\begin{figure}[htbp]
\centering
\includegraphics[width=\textwidth]{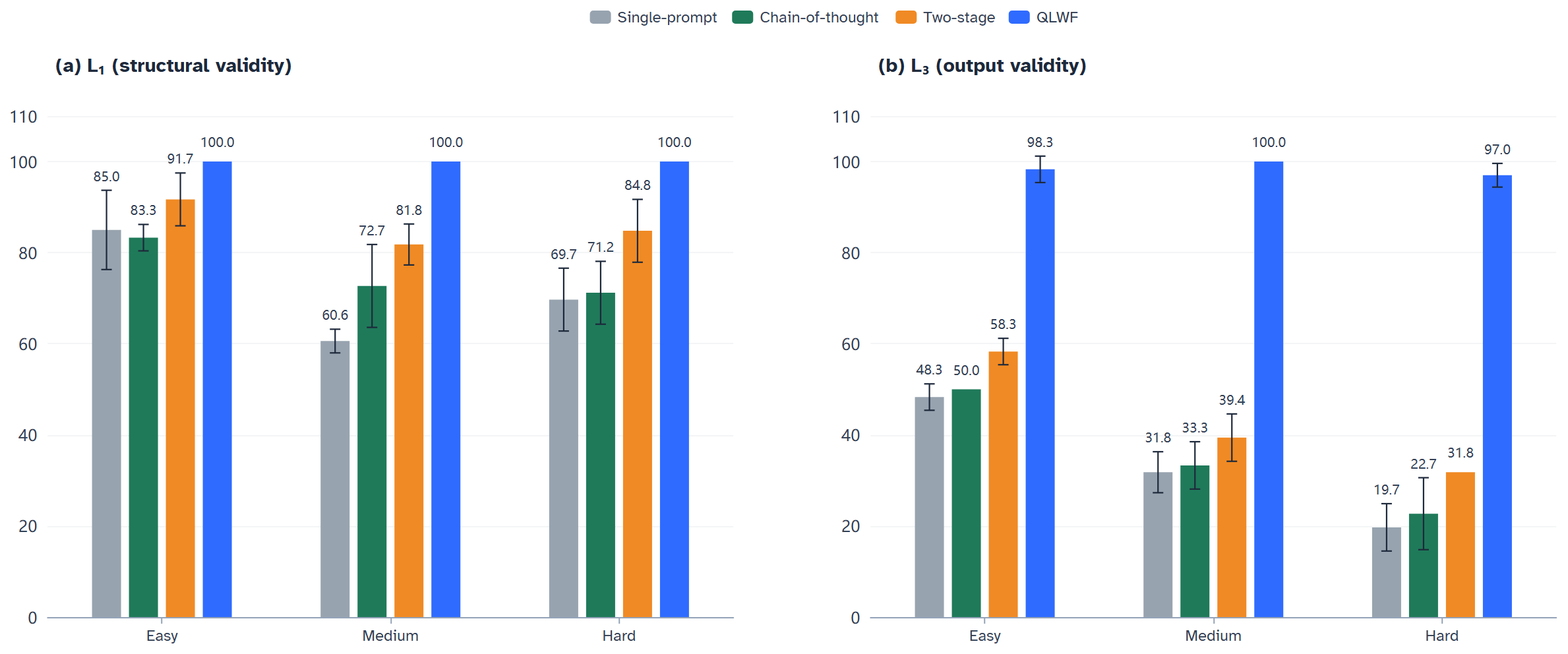}
\caption{Pass rates by difficulty subgroup across four generation strategies. (a) $L_1$ structural validity, (b) $L_3$ output validity. Error bars show $\pm$1 SD across 3 independent runs. QLWF remains stable across difficulty levels, while prompt-based baselines decline sharply on harder tasks.}\label{fig:difficulty}
\end{figure}

Figure~\ref{fig:cascade} provides a cascade view of this progressive filtering. Table~\ref{tab:difficulty} breaks down the results by difficulty level. QLWF maintains $L_1$ = 100\% and $L_3 \geq 97.0$\% across all three levels (Easy, Medium, and Hard). The prompt-based baselines show a different pattern, though: their pass rates decline sharply as difficulty increases. For example, Single-prompt drops from 48.3\% $L_3$ on Easy tasks to 19.7\% on Hard tasks. Figure~\ref{fig:difficulty} illustrates this divergence. The stability of QLWF across difficulty levels suggests that the structured pipeline is not limited to simple tasks but scales to more complex analytical requirements within the current benchmark.

\begin{table}[htbp]
\caption{QL-Bench results by difficulty subgroup. $L_1$ and $L_3$ pass rates (\%) are reported for each subset. Values are means across 3 independent runs.}\label{tab:difficulty}
\begin{tabular*}{\textwidth}{@{\extracolsep\fill}lrrrrrr}
\toprule
& \multicolumn{2}{c}{Easy ($n = 20$)} & \multicolumn{2}{c}{Medium ($n = 22$)} & \multicolumn{2}{c}{Hard ($n = 22$)} \\
\cmidrule(lr){2-3}\cmidrule(lr){4-5}\cmidrule(lr){6-7}
Method & $L_1$ & $L_3$ & $L_1$ & $L_3$ & $L_1$ & $L_3$ \\
\midrule
Single-prompt & 85.0 & 48.3 & 60.6 & 31.8 & 69.7 & 19.7 \\
Chain-of-thought & 83.3 & 50.0 & 72.7 & 33.3 & 71.2 & 22.7 \\
Two-stage & 91.7 & 58.3 & 81.8 & 39.4 & 84.8 & 31.8 \\
QLWF & 100.0 & 98.3 & 100.0 & 100.0 & 100.0 & 97.0 \\
\bottomrule
\end{tabular*}
\end{table}

Table~\ref{tab:by-direction} breaks down $L_3$ pass rates by research direction across all four strategies. QLWF achieves full $L_3$ compliance in eight of the ten directions, with dependency direction at 94.4\% and word length at 88.9\%. The two shortfalls reflect occasional node-configuration challenges in tasks that require directional dependency counts or character-level length distributions. The prompt-based baselines show far more variable coverage. Entropy, Lexical Diversity, and Word-Length Frequency collapse to zero under Single-prompt, and several directions reach at most 50--75\% under Two-stage. We report QLWF's two sub-100\% directions rather than suppress them, as they help define the current boundaries of the approach.

\begin{table}[htbp]
\caption{QL-Bench $L_3$ pass rates (\%) by research direction across four generation strategies. Values are 3-run means. Overall pass rates are reported in Table~\ref{tab:main-results}.}\label{tab:by-direction}
\begin{tabular*}{\textwidth}{@{\extracolsep\fill}lrrrrr}
\toprule
Research direction & Tasks & Single-prompt & Chain-of-thought & Two-stage & QLWF \\
\midrule
Dependency direction   & 6 & 66.7 & 50.0 & 66.7 & 94.4 \\
Dependency distance    & 8 & 37.5 & 75.0 & 87.5 & 100.0 \\
Entropy                & 6 & 0.0  & 0.0  & 16.7 & 100.0 \\
Lexical diversity      & 6 & 0.0  & 33.3 & 0.0  & 100.0 \\
Syntactic network      & 8 & 62.5 & 50.0 & 75.0 & 100.0 \\
Tree metrics           & 6 & 33.3 & 83.3 & 66.7 & 100.0 \\
Valency                & 6 & 50.0 & 50.0 & 50.0 & 100.0 \\
Word length            & 6 & 16.7 & 0.0  & 16.7 & 88.9  \\
Word-length frequency  & 6 & 0.0  & 16.7 & 0.0  & 100.0 \\
Zipf law               & 6 & 33.3 & 0.0  & 0.0  & 100.0 \\
\bottomrule
\end{tabular*}
\end{table}

\begin{figure}[htbp]
\centering
\includegraphics[width=\textwidth]{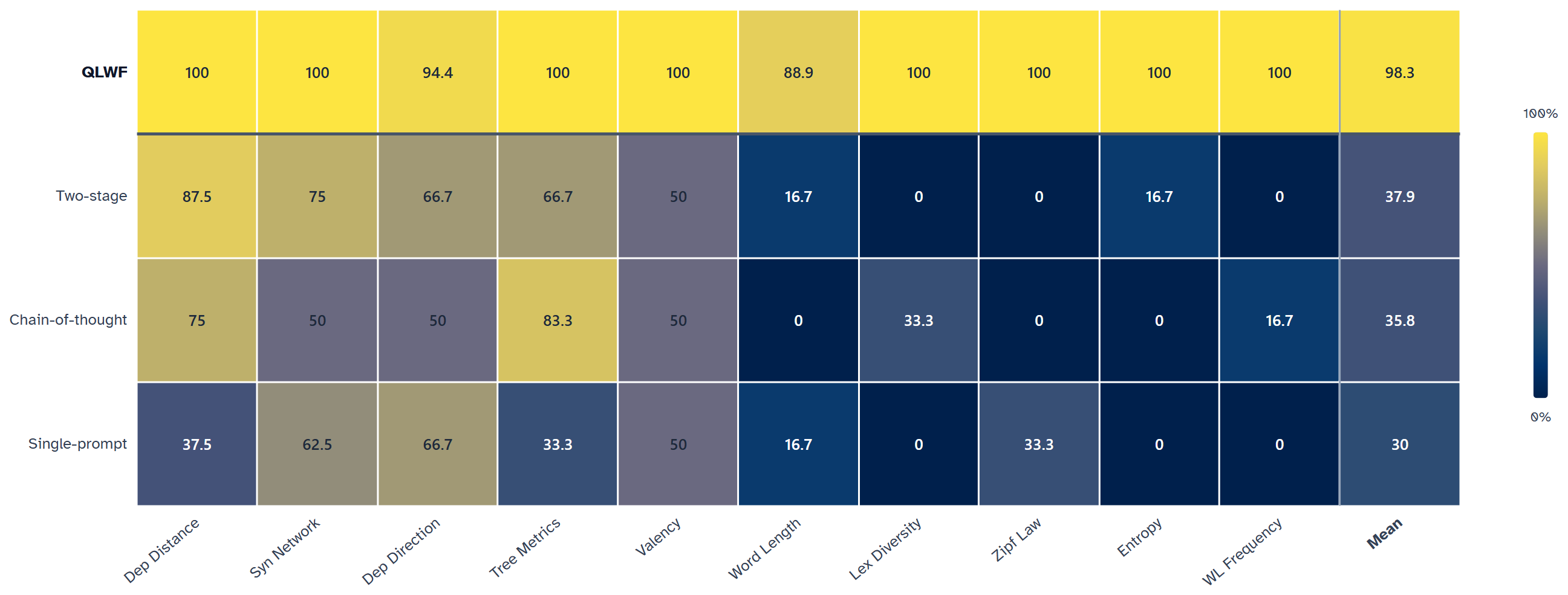}
\caption{$L_3$ pass rates (\%) by research direction and generation strategy, rendered with a sequential cividis palette (dark = low, light = high). Columns are ordered left-to-right by baseline mean pass rate, so directions the three prompt-based methods handle relatively well appear on the left and those they struggle on appear on the right. The rightmost column reports each method's mean across all ten directions. QLWF maintains near-universal compliance across directions (mean $98.3$\%), while the three prompt-based baselines show increasingly uneven coverage as one scans rightward (means $30.0$--$37.9$\%).}\label{fig:heatmap}
\end{figure}

Figure~\ref{fig:heatmap} presents the same data as a heatmap, visually emphasising the contrast between QLWF's near-uniform coverage and the patchy performance of the three baselines.

\subsection{Incremental Refinement and Workflow Lifecycle}\label{sec:refinement-results}

\begin{figure}[htbp]
\centering
\includegraphics[width=\textwidth]{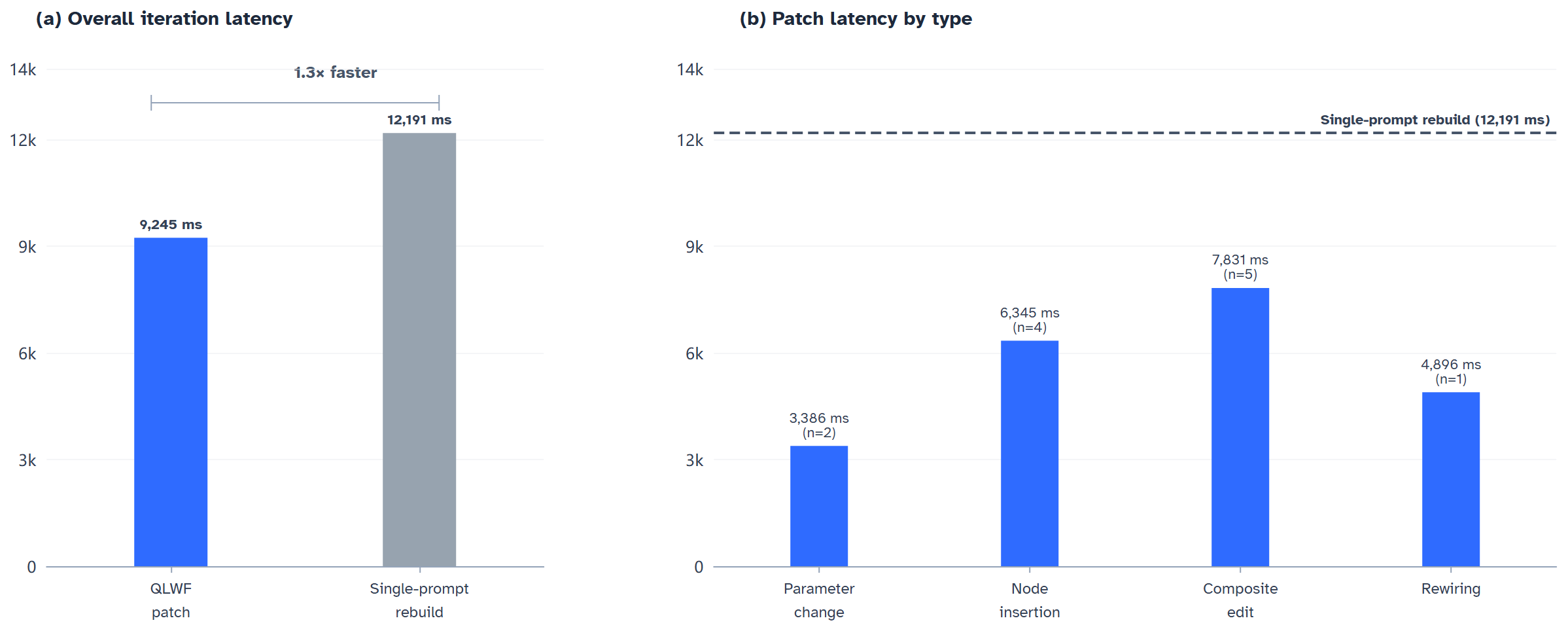}
\caption{Lifecycle benchmark results. (a) Mean iteration latency for QLWF patch versus Single-prompt rebuild, pooled across $3 \times 12 = 36$ trials. (b) Mean latency of the QLWF patch operation broken down by edit type, with category $n$ marking the task count. The dashed reference line in (b) marks the Single-prompt rebuild mean ($12{,}191$~ms) from (a). All four patch-type means fall well below it. Both panels share a $0$--$14$~k~ms y-axis for direct height comparison.}\label{fig:lifecycle}
\end{figure}

Table~\ref{tab:lifecycle} compares QLWF patch-based refinement with Single-prompt full regeneration on the 12-task lifecycle benchmark. Across three runs (36 trials in total), QLWF patch succeeds in all 36 cases (100\%), while Single-prompt rebuild succeeds in 35 out of 36 (97.2\%). The cost difference is more pronounced: QLWF patch uses a mean of 14,855 total tokens per run ($\pm$ 3,222) versus 47,187 ($\pm$ 2,443) for Single-prompt rebuild, and achieves a mean latency of 9,245~ms ($\pm$ 3,017) versus 12,191~ms ($\pm$ 1,505). Figure~\ref{fig:lifecycle}(a) visualises the mean latencies side by side. Latency was measured under parallel API execution and should be interpreted as approximate. These results indicate that, within this matched benchmark, incremental refinement is both more reliable and, in terms of token cost, substantially more efficient than full reconstruction.

\begin{table}[htbp]
\caption{Lifecycle iteration comparison between QLWF patch editing and single-prompt rebuild on a matched 12-task benchmark. Success pooled across $3 \times 12 = 36$ trials. Tokens and latency are mean $\pm$ std across 3 runs. Latency is approximate (parallel execution).}\label{tab:lifecycle}
\begin{tabular*}{\textwidth}{@{\extracolsep\fill}lrrr@{\,$\pm$\,}lr@{\,$\pm$\,}l}
\toprule
Method & Trials & Success rate (\%) & \multicolumn{2}{c}{Total tokens} & \multicolumn{2}{c}{Mean latency (ms)} \\
\midrule
QLWF patch           & 36 & 100.0 (36/36) & 14{,}855 & 3{,}222 & 9{,}245  & 3{,}017 \\
Single-prompt rebuild & 36 & 97.2 (35/36) & 47{,}187 & 2{,}443 & 12{,}191 & 1{,}505 \\
SP / QLWF ratio      & -- & --            & \multicolumn{2}{c}{3.2$\times$} & \multicolumn{2}{c}{--} \\
\bottomrule
\end{tabular*}
\end{table}

Beyond overall success, three fidelity metrics characterise patch quality. The first-attempt success rate (Success@0) is 100\%, the fallback rate is 0\%, and the unrelated-change rate is 0\%. The mean patch size is 3.6 operations per workflow. These numbers indicate that the refinement operations are not only successful but precise: each patch modifies exactly the intended part of the workflow without introducing unrelated changes. This structural preservation is the core methodological advantage of incremental refinement: rather than discarding a validated workflow and regenerating it from scratch, the researcher modifies only the parts that need to change, while the rest of the verified structure remains intact. These metrics are consistent across all three runs.

All four edit categories (parameter change, node insertion, composite edit, and rewiring) achieve 100\% success. Latency varies by category: parameter changes have the lowest latency (mean 3,386~ms across three runs), followed by rewiring (4,896~ms), node insertions (6,345~ms), and composite edits (7,831~ms). Figure~\ref{fig:lifecycle}(b) shows the per-category mean latency against the Single-prompt rebuild reference from panel (a).

\begin{figure}[htbp]
\centering
\includegraphics[width=\textwidth]{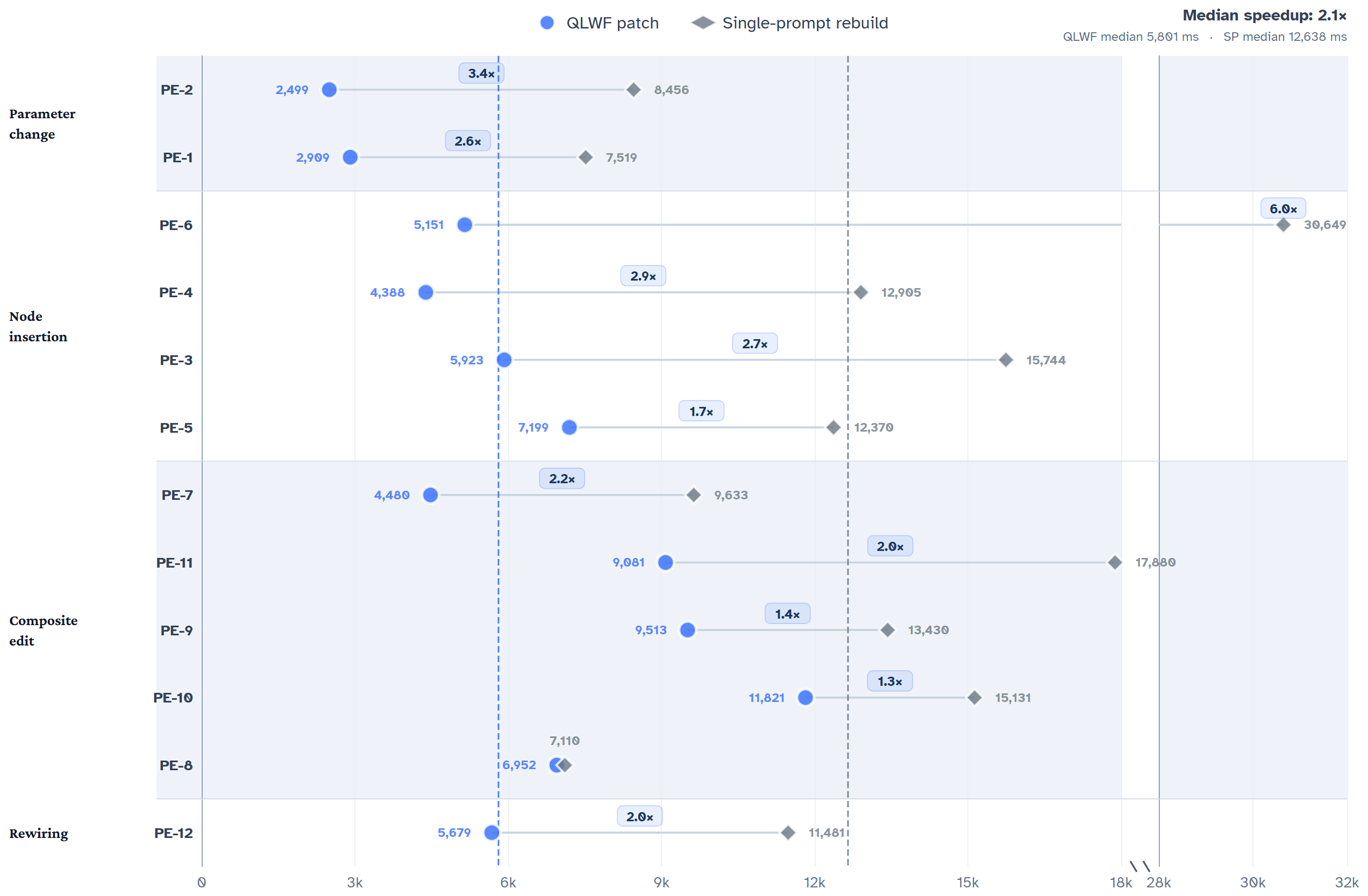}
\caption{Per-task latency comparison between QLWF patch (circles) and Single-prompt rebuild (diamonds) across the 12 lifecycle benchmark tasks (Run~0 detail). Rows are grouped by edit category, and each connector line is annotated with the row's SP-to-QLWF speedup ratio. Vertical dashed lines mark the per-method medians ($5{,}801$ ms for QLWF, $12{,}638$ ms for SP). The x-axis is broken between $20$~k and $28$~k~ms to accommodate the PE-6 Single-prompt outlier ($30{,}649$~ms) without compressing the other eleven tasks. A double slash (//) on the axis marks the break.}\label{fig:paired-lifecycle}
\end{figure}

Figure~\ref{fig:paired-lifecycle} shows the per-task latency comparison. The 12-task benchmark (36 trials across three runs) is a hard constraint on this dimension of the evaluation. The results cannot be extrapolated to general-purpose editing capabilities. Our conclusion is limited to the matched local benchmark: within this controlled setting, patch-based refinement is feasible and precise.

\subsection{Methodological Cost and Boundary Conditions}\label{sec:cost}

The cost profile across both experimental dimensions can be read directly from Tables~\ref{tab:main-results} and~\ref{tab:lifecycle}. For the 64-task construction benchmark, QLWF uses approximately 1.7 times the tokens of the Single-prompt baseline, with a $+65.6$ pp gain in $L_3$ pass rate. For the 12-task refinement benchmark, QLWF patch uses approximately one-third of the tokens of Single-prompt rebuild while matching or exceeding its success rate. Token counts throughout this study are approximate values derived from the evaluation runner's stage-level aggregation. They should be treated as indicative cost estimates, not as precise billing figures.

We also conducted an ablation on Progressive Node Disclosure (PND), a mechanism that incrementally reveals node candidates to the language model during the select stage rather than presenting the full library at once (Table~\ref{tab:pnd-ablation}). On the 64-task benchmark, PND reduces total token consumption by 76.7\% (from 2,564,953 to 596,847 tokens) while maintaining the same structure-level $\text{Pass@1}$ rate (96.9\%). This result demonstrates that PND is an effective cost-reduction mechanism at the structural level. This ablation is limited to structure-level $\text{Pass@1}$ and should not be interpreted as a statement about $L_3$ output plausibility or overall workflow quality.

\begin{table}[htbp]
\caption{PND ablation on the 64-task benchmark (structure-level $\text{Pass@1}$ via expected-node coverage).}\label{tab:pnd-ablation}
\begin{tabular*}{\textwidth}{@{\extracolsep\fill}lrrr}
\toprule
Setting & Structure-level $\text{Pass@1}$ (\%) & Total tokens & Token reduction (\%) \\
\midrule
QLWF w/o PND & 96.9 & 2,564,953 & -- \\
QLWF + PND & 96.9 & 596,847 & 76.7 \\
\bottomrule
\end{tabular*}
\end{table}


\FloatBarrier
\section{Evaluation and Discussion}\label{sec:discussion}

\subsection{Methodological Implications}\label{sec:implications}

The experimental results reported in Section~\ref{sec:experiments} corroborate the reification-formalization framework introduced in Section~\ref{sec:workflow-object}. The construction experiments (RQ1) demonstrate that a researcher's natural-language description of an analytical task can be automatically reified into a visual, inspectable workflow structure, and that this structure can be formalised into a deterministic, executable data pipeline. Across three runs on the 64-task QL-Bench benchmark, the proposed pipeline achieves full structural and executability compliance with a mean output-plausibility rate of 98.4\%. Concretely, a workflow produced by the pipeline can be handed to another researcher with reasonable confidence that it will run correctly and yield valid results. The refinement experiments (RQ2) show that the resulting workflow objects, once saved, can be incrementally revised through targeted patches without sacrificing structural integrity. For a research artefact meant to be reused over time, this matters: the workflow can evolve with the researcher's needs rather than be discarded and rebuilt. Collectively, these results indicate that the dual transformation of reification and formalization is a viable methodological path for organising quantitative-syntax research procedures. The language model's role, however, remains limited to the construction stage. The researcher decides whether a generated or revised workflow correctly captures the intended analytical logic, and whether the outputs carry linguistic meaning.

The four generation strategies compared in this study form a progression of increasing structural commitment that helps explain why the pipeline approach is effective. Single-prompt imposes no intermediate structure. Chain-of-thought introduces implicit reasoning within a single response. Two-stage creates an explicit intermediate plan in a separate call. QLWF passes the input through five constrained stages, each injecting domain-specific context. Output validity improves monotonically along this progression, which suggests the gain comes less from better prompting alone and more from how far the generation process has been formalised into structured, domain-aware stages. The quality of the generated workflow, then, depends on the quality of the domain resources it draws on: the node library, the type contracts, and the validation rules. These resources are themselves shareable and extensible.

The closest point of comparison in the LRAE literature is TextFlows \cite{perovsek2016,martinc2024}, which follows a similar overall logic: a platform is presented, applied to a bounded use case, evaluated experimentally, and discussed with restrained conclusions. Our work shares this structure but extends it along three dimensions. First, QLWF adds AI-assisted construction, allowing workflows to be generated from natural-language descriptions rather than assembled manually. Second, it introduces an incremental refinement mechanism that supports targeted revision of saved workflow objects. Third, it provides an explicit theoretical framing through the reification-formalization lens, which positions the workflow as a reusable language resource rather than a software artefact alone. Such a resource can be shared across research groups, evaluated under controlled conditions, and extended to new analytical tasks.

\subsection{Language Resources and Evaluation Potential}\label{sec:resources}

Corpora, lexicons, and annotation standards are valued as language resources because they can be accessed, reused, and evaluated by the research community. We argue that workflow objects share these properties, and the present study contributes three such resources to the quantitative-syntax community. The first is the domain-specific node library, comprising 48 functional processing nodes organised into nine categories. These nodes encode the core operations of quantitative-syntax research and can be used directly by other researchers to compose new analytical workflows without reimplementing standard metrics. The second is the QL-Bench benchmark suite, consisting of 64 task definitions in a structured JSONL format, each with a natural-language description, expected node combination, and predefined $L_3$ validation rules. Other research teams can adopt this benchmark as a testbed or extend it with additional tasks. The third is a set of six workflow templates covering common quantitative-syntax analysis scenarios, from basic dependency metrics to full multi-indicator pipelines. These templates are exportable as self-contained JSON files and can serve as starting points for new studies. All three resource types are integrated into QLWF, a fully functional platform that researchers can install and use to construct, execute, share, and revise workflows. The platform is itself a resource. As the usage scenario in Section~\ref{sec:usage-scenario} illustrates, a single researcher can move from natural-language intent to a shareable workflow object within one session, and can do so without programming expertise. In the language-resource tradition, software that makes reuse possible is itself counted as a resource. In that sense, QLWF itself belongs on the list as a fourth resource: an entry point for the community to construct, inspect, and reuse quantitative-syntax workflows. The node categorisation scheme and the QL-Bench task taxonomy can also function as a lightweight domain ontology for quantitative syntax. The node categories encode the operational space of the field, while the task directions encode its problem space. Formalising these structures into a full ontology is left for future work.

Beyond these resources, the reification-formalization framework also has implications for evaluation. When four generation strategies are compared on the same benchmark under the same conditions, the workflow object itself becomes the unit of comparison, which means that the three-level assessment protocol functions as an evaluation instrument: $L_1$ measures whether research intent was correctly reified into a node-flow structure, $L_2$ measures whether the structure has valid execution semantics, and $L_3$ measures whether the end-to-end pipeline produces plausible output. The situation parallels the evaluation of corpora or annotation standards in the language-resource tradition, where the object itself is the focus of assessment. This comparison illustrates the potential of workflow-based evaluation as a general approach to assessing language-research procedures. Central to this potential is the objectivity of the $L_3$ assessment. Because all acceptance thresholds are derived from published quantitative-syntax results and fixed before the experiment begins, the evaluation does not depend on post-hoc human judgement. This pre-registered, automated design is what makes the comparison across strategies fair and reproducible. The three-level structure is itself transferable: researchers in other subfields could adopt the same logic by supplying their own validation criteria.

\subsection{Generalizability and Limitations}\label{sec:limitations}

How far can these results reach? Generalisation here does not mean stripping out domain specificity. It means redoing the depth-configuration approach in another field. The experiments in Section~\ref{sec:experiments} show that when domain-specific depth is configured well enough (a dedicated node library, a literature-grounded benchmark, and automated output validation), the workflow-centred approach is highly effective within its bounded case. The open question is whether the same configuration can be reproduced in other language-research subfields. Within quantitative linguistics, lexical statistics, Zipf-law modelling across genres, and language-diversity measurement are natural next candidates. Beyond quantitative linguistics, research workflows in computational stylistics, translation studies, and corpus-based discourse analysis could also benefit from this route. Each subfield would need its own node library, its own benchmark tasks, and its own output validation criteria. Once each is in place, the outcome in each case is a set of shareable language resources tailored to that community.

One useful point of comparison is InstructPipe \cite{zhou2025}, which targets general-purpose ML pipelines and does not specialise to a domain. InstructPipe shows that AI-assisted pipeline construction is broadly feasible across domains. Our study fills in the other side: within a single domain, configuring domain-specific depth well can sharply improve output validity. Breadth and depth are two sides of the same coin. Researchers who want to replicate this depth-first path in a new language-research subfield need three things in place: (1)~a domain-specific node library that encodes the field's core operations, (2)~a domain-grounded benchmark task set with predefined validation criteria, and (3)~domain-specific output validation standards derived from the field's established results. Each of these is itself a reusable resource. Together, they form the foundation on which a community can build, share, and evaluate analytical workflows.

This study has clear limits. On internal validity, language-model outputs are inherently non-deterministic, and the three-run design can reduce run-to-run variance but cannot remove it. The latency measurements were also collected under parallel API execution and should be read as approximate. On external validity, QL-Bench is a researcher-built local benchmark, not a community-endorsed standard. The patch benchmark covers only 12 tasks (36 trials across three runs), and all conclusions remain bounded by the quantitative-syntax domain. On construct validity, the $L_3$ protocol is scoped to computational output plausibility, and the linguistic interpretation of the results remains, by design, the researcher's responsibility. Token counts are approximate values from runner-level aggregation, not vendor billing figures. The workflows we produce are deterministic data-flow pipelines, not agent workflows, and the two should not be conflated. All experiments use the same GLM-5 backend introduced in Section~\ref{sec:setup}. Because the language model participates only at design time through typed stage contracts, the pipeline is architecturally model-agnostic, and replicating the results on other LLMs is primarily an engineering step.

The discussion so far has centred on analytical workflows, the workflows that process existing linguistic data. The same paradigm can also support the construction of language resources. A workflow that chains tokenisation, part-of-speech tagging, and manual validation, for example, can serve as a reproducible corpus-annotation pipeline. Workflows for lexicon compilation or annotation-scheme comparison benefit from the same combination of visual inspectability and deterministic execution. To support this broader potential, we plan to release the QLWF platform as open-source software in the near future. The release will include the node library, QL-Bench benchmark, and workflow templates described in this paper. With the release, research logic becomes visible and editable instead of being buried in ad-hoc scripts, which can accelerate independent verification and support the collaborative extension of language-research workflows.


\section{Conclusion}\label{sec:conclusion}

The analytical procedures that underpin quantitative language research need to be organised as language resources in their own right: objects that can be inspected, shared, and revised. This paper has framed this need through the dual lens of reification and formalization: research logic is first externalised as a visible workflow structure, then given deterministic execution semantics.

Three research questions were addressed within the bounded domain of quantitative syntax. For RQ1, the construction experiments showed that natural-language research descriptions can be transformed into executable workflows through a five-stage, AI-assisted pipeline. Across three runs on a 64-task benchmark, the pipeline achieves full structural and executability compliance with a mean output-plausibility rate of 98.4\%. For RQ2, the refinement experiments showed that saved workflow objects can be incrementally revised through targeted patches. Across 36 trials, patch-based refinement succeeds in all cases with zero unrelated changes, while using approximately one-third of the tokens required by full regeneration. For RQ3, the methodological cost of the structured pipeline is moderate (a token ratio of approximately 1.7 relative to the simplest baseline), and all conclusions remain bounded by the current benchmark and domain.

Beyond these methodological findings, the study contributes a set of reusable language-research resources: a domain-specific node library (48 nodes across nine categories), a literature-grounded benchmark suite (QL-Bench, 64 tasks), six workflow templates, and the QLWF platform through which these resources can be accessed and used. The three-level assessment protocol ($L_1$, $L_2$, $L_3$) also illustrates the potential of a workflow-based evaluation framework for comparing generation strategies under controlled conditions.

Three directions for future work emerge from this study. The first is to replicate the depth-configuration approach in other language-research subfields, both within quantitative linguistics and beyond, each requiring its own node library, benchmark, and validation criteria. A related direction is to extend the workflow paradigm from analysis to the construction of language resources such as annotated corpora and lexicons. The second is to move towards community standardisation of QL-Bench by inviting other researchers to contribute tasks and validation rules, gradually transforming it from a local benchmark into a shared resource. The planned open-source release of QLWF is intended as a first step in this direction. The third is to develop the node categorisation scheme and the task taxonomy into a formal domain ontology for quantitative syntax, which could be aligned with existing language ontologies such as GOLD and OLIA.


\backmatter

\section*{Declarations}

\bmhead{Funding}
This work was supported by the National Social Science Fund of China under the project ``Research on the Construction and Application of a Generative AI-Enabled Multimodal Disciplinary Knowledge Graph for Linguistics'' (Grant No.~24BYY079).

\bmhead{Competing interests}
The authors declare no competing interests.

\bmhead{Data availability}
The workflow-generation benchmark, lifecycle patch benchmark, workflow templates, and repository-managed runtime resources are maintained in a private GitHub repository during peer review and are available from the corresponding author on reasonable request. The underlying Universal Dependencies corpora are publicly available from the Universal Dependencies project, and the pinned source metadata used in this study are recorded in the repository manifest.

\bmhead{Code availability}
The QLWF platform code is maintained in a private GitHub repository during peer review and is available from the corresponding author on reasonable request. A public repository release is planned upon publication.

\bmhead{Author contributions}
He Wang: methodology, software, experiments, and writing. Jingbo Chen: resources, validation, and review. Yuqiao Lai: data curation and review. Nan Yang: verification of operators and metrics. Hanwen Zhang: verification of figure and table details. Wei Yuan: supervision, conceptualization, corresponding author responsibilities, and review.

\bibliography{references}

@book{wenger1998,
  author    = {Wenger, Etienne},
  title     = {Communities of Practice: Learning, Meaning, and Identity},
  publisher = {Cambridge University Press},
  year      = {1998},
  address   = {Cambridge}
}

@article{montague1970,
  author  = {Montague, Richard},
  title   = {Universal grammar},
  journal = {Theoria},
  year    = {1970},
  volume  = {36},
  number  = {3},
  pages   = {373--398},
  doi     = {10.1111/j.1755-2567.1970.tb00434.x}
}

@article{liu2008,
  author  = {Liu, Haitao},
  title   = {Dependency distance as a metric of language comprehension difficulty},
  journal = {Journal of Cognitive Science},
  year    = {2008},
  volume  = {9},
  number  = {2},
  pages   = {159--191},
  doi     = {10.17791/jcs.2008.9.2.159}
}

@article{lu2010,
  author  = {Lu, Xiaofei},
  title   = {Automatic analysis of syntactic complexity in second language writing},
  journal = {International Journal of Corpus Linguistics},
  year    = {2010},
  volume  = {15},
  number  = {4},
  pages   = {474--496},
  doi     = {10.1075/ijcl.15.4.02lu}
}

@book{zipf1949,
  author    = {Zipf, George Kingsley},
  title     = {Human Behavior and the Principle of Least Effort},
  publisher = {Addison-Wesley},
  year      = {1949},
  address   = {Cambridge, MA}
}

@article{perovsek2016,
  author  = {Perov\v{s}ek, Matic and Kranjc, Janez and Erjavec, Toma\v{z} and Cestnik, Bojan and Lavra\v{c}, Nada},
  title   = {{TextFlows}: A visual programming platform for text mining and natural language processing},
  journal = {Science of Computer Programming},
  year    = {2016},
  volume  = {121},
  pages   = {128--152},
  doi     = {10.1016/j.scico.2016.01.001}
}

@article{martinc2024,
  author  = {Martinc, Matej and Perov\v{s}ek, Matic and Lavra\v{c}, Nada and Pollak, Senja},
  title   = {{TextFlows}: An open science {NLP} evaluation approach},
  journal = {Language Resources and Evaluation},
  year    = {2024},
  volume  = {59},
  pages   = {4439--4468},
  doi     = {10.1007/s10579-024-09793-1}
}

@inproceedings{hinrichs2010,
  author    = {Hinrichs, Marie and Zastrow, Thomas and Hinrichs, Erhard},
  title     = {{WebLicht}: Web-based {LRT} services in a distributed {eScience} infrastructure},
  booktitle = {Proceedings of the Seventh International Conference on Language Resources and Evaluation ({LREC} 2010)},
  year      = {2010},
  publisher = {European Language Resources Association},
  address   = {Valletta, Malta},
  url       = {https://aclanthology.org/L10-1184/}
}

@article{cunningham2002,
  author  = {Cunningham, Hamish},
  title   = {{GATE}, a general architecture for text engineering},
  journal = {Computers and the Humanities},
  year    = {2002},
  volume  = {36},
  number  = {2},
  pages   = {223--254},
  doi     = {10.1023/A:1014348124664}
}

@inproceedings{ide2014,
  author    = {Ide, Nancy and Pustejovsky, James and Cieri, Christopher and Nyberg, Eric and Wang, Di and Suderman, Keith and Verhagen, Marc and Wright, Jonathan},
  title     = {The {Language Application Grid}},
  booktitle = {Proceedings of the Ninth International Conference on Language Resources and Evaluation ({LREC} 2014)},
  year      = {2014},
  publisher = {European Language Resources Association},
  address   = {Reykjavik, Iceland},
  pages     = {22--30},
  url       = {https://aclanthology.org/L14-1706/}
}

@inproceedings{labropoulou2018,
  author    = {Labropoulou, Penny and Galanis, Dimitrios and Lempesis, Antonis and Greenwood, Mark and Knoth, Petr and Eckart de Castilho, Richard and Sachtouris, Stavros and Georgantopoulos, Byron and Anastasiou, Lucas and Martziou, Stefania and Gkirtzou, Katerina and Manola, Natalia and Piperidis, Stelios},
  title     = {{OpenMinTeD}: A platform facilitating text mining of scholarly content},
  booktitle = {WOSP 2018 Workshop Proceedings},
  year      = {2018},
  publisher = {European Language Resources Association},
  address   = {Luxemburg},
  pages     = {7--12}
}

@article{berthold2009,
  author  = {Berthold, Michael R. and Cebron, Nicolas and Dill, Fabian and Gabriel, Thomas R. and K\"{o}tter, Tobias and Meinl, Thorsten and Ong, Peter and Sieb, Christoph and Thiel, Kilian and Wiswedel, Bernd},
  title   = {{KNIME}---the {Konstanz} information miner: version 2.0 and beyond},
  journal = {ACM SIGKDD Explorations Newsletter},
  year    = {2009},
  volume  = {11},
  number  = {1},
  pages   = {26--31},
  doi     = {10.1145/1656274.1656280}
}

@article{gomes2025,
  author  = {Gomes, Lu{\'i}s and Branco, Ant{\'o}nio and Silva, Jo{\~a}o and Branco, Ruben},
  title   = {From greatest simplicity to full power: language technology infrastructures and the path to research-infrastructure-as-a-service for multiple user groups},
  journal = {Language Resources and Evaluation},
  year    = {2025},
  volume  = {59},
  pages   = {4391--4420},
  doi     = {10.1007/s10579-024-09772-6}
}

@article{repar2020,
  author  = {Repar, Andra{\v{z}} and Pollak, Senja and Kranjc, Janez},
  title   = {Reproduction, replication, analysis and adaptation of a term alignment approach},
  journal = {Language Resources and Evaluation},
  year    = {2020},
  volume  = {54},
  pages   = {767--800},
  doi     = {10.1007/s10579-019-09477-1}
}

@inproceedings{zhou2025,
  author    = {Zhou, Zhongyi and Jin, Jing and Phadnis, Vrushank and Yuan, Xiuxiu and Jiang, Jun and Qian, Xun and Wright, Kristen and Sherwood, Mark and Mayes, Jason and Zhou, Jingtao and Huang, Yiyi and Xu, Zheng and Zhang, Yinda and Lee, Johnny and Olwal, Alex and Kim, David and Iyengar, Ram and Li, Na and Du, Ruofei},
  title     = {{InstructPipe}: Generating visual blocks pipelines with human instructions and {LLMs}},
  booktitle = {Proceedings of the 2025 {CHI} Conference on Human Factors in Computing Systems},
  year      = {2025},
  publisher = {Association for Computing Machinery},
  address   = {New York, NY, USA},
  pages     = {1--22},
  doi       = {10.1145/3706598.3713905}
}

@inproceedings{zhang2025,
  author    = {Zhang, Jiayi and Xiang, Jinyu and Yu, Zhaoyang and Teng, Fengwei and Chen, Xiong-Hui and Chen, Jiaqi and Zhuge, Mingchen and Cheng, Xin and Hong, Sirui and Wang, Jinlin and Zheng, Bingnan and Liu, Bang and Luo, Yuyu and Wu, Chenglin},
  title     = {{AFlow}: Automating agentic workflow generation},
  booktitle = {The Thirteenth International Conference on Learning Representations},
  year      = {2025},
  address   = {Singapore},
  url       = {https://openreview.net/forum?id=z5uVAKwmjf}
}

@inproceedings{tan2025,
  author    = {Tan, Xiaoyu and Li, Bin and Qiu, Xihe and Qu, Chao and Chu, Wei and Xu, Yinghui and Qi, Yuan},
  title     = {{Meta-Agent-Workflow}: Streamlining tool usage in {LLMs} through workflow construction, retrieval, and refinement},
  booktitle = {Companion Proceedings of the {ACM} on Web Conference 2025},
  year      = {2025},
  publisher = {Association for Computing Machinery},
  address   = {New York, NY, USA},
  pages     = {458--467},
  doi       = {10.1145/3701716.3715247}
}

@inproceedings{yin2023,
  author    = {Yin, Pengcheng and Neubig, Graham and Yao, Shunyu and others},
  title     = {Natural language to code generation in interactive data science notebooks},
  booktitle = {Proceedings of the 2023 Conference on Empirical Methods in Natural Language Processing},
  year      = {2023},
  publisher = {Association for Computational Linguistics},
  address   = {Singapore},
  pages     = {12553--12574},
  doi       = {10.18653/v1/2023.emnlp-main.774},
  url       = {https://aclanthology.org/2023.emnlp-main.774/}
}

@inproceedings{du2022,
  author    = {Du, Zhengxiao and Qian, Yujie and Liu, Xiao and others},
  title     = {{GLM}: General language model pretraining with autoregressive blank infilling},
  booktitle = {Proceedings of the 60th Annual Meeting of the Association for Computational Linguistics (Volume 1: Long Papers)},
  year      = {2022},
  publisher = {Association for Computational Linguistics},
  address   = {Dublin, Ireland},
  pages     = {320--335},
  doi       = {10.18653/v1/2022.acl-long.26},
  url       = {https://aclanthology.org/2022.acl-long.26/}
}

@article{crusoe2022,
  author  = {Crusoe, Michael R. and Abeln, Sanne and Iosup, Alexandru and Amstutz, Peter and Chilton, John and Tijani\'{c}, Neboj\v{s}a and M\'{e}nager, Herv\'{e} and Soiland-Reyes, Stian},
  title   = {Methods included: Standardizing computational reuse and portability with the {Common Workflow Language}},
  journal = {Communications of the ACM},
  year    = {2022},
  volume  = {65},
  number  = {6},
  pages   = {54--63},
  doi     = {10.1145/3486897}
}

@article{goble2020,
  author  = {Goble, Carole and Cohen-Boulakia, Sarah and Soiland-Reyes, Stian and Garijo, Daniel and Gil, Yolanda and Crusoe, Michael R. and Peters, Kristian and Schober, Daniel},
  title   = {{FAIR} computational workflows},
  journal = {Data Intelligence},
  year    = {2020},
  volume  = {2},
  number  = {1--2},
  pages   = {108--121},
  doi     = {10.1162/dint_a_00033}
}

@article{gustafsson2025,
  author  = {Gustafsson, Ove Johan Ragnar and Wilkinson, Sean R. and Bacall, Finn and Soiland-Reyes, Stian and Leo, Simone and Pireddu, Luca and Owen, Stuart and Juty, Nick and Fern\'{a}ndez, Jos\'{e} M. and Brown, Tom and M\'{e}nager, Herv\'{e} and Gr{\"u}ning, Bj{\"o}rn and Capella-Gutierrez, Salvador and Coppens, Frederik and Goble, Carole},
  title   = {{WorkflowHub}: A registry for computational workflows},
  journal = {Scientific Data},
  year    = {2025},
  volume  = {12},
  number  = {1},
  pages   = {837},
  doi     = {10.1038/s41597-025-04786-3}
}

@article{ewels2020,
  author  = {Ewels, Philip A. and Peltzer, Alexander and Fillinger, Sven and Patel, Harshil and Alneberg, Johannes and Wilm, Andreas and Garcia, Maxime Ulysse and Di Tommaso, Paolo and Nahnsen, Sven},
  title   = {The nf-core framework for community-curated bioinformatics pipelines},
  journal = {Nature Biotechnology},
  year    = {2020},
  volume  = {38},
  number  = {3},
  pages   = {276--278},
  doi     = {10.1038/s41587-020-0439-x}
}

@article{missier2016,
  author  = {Missier, Paolo},
  title   = {Provenance and data differencing for workflow reproducibility analysis},
  journal = {Concurrency and Computation: Practice and Experience},
  year    = {2016},
  volume  = {28},
  number  = {4},
  pages   = {995--1015},
  doi     = {10.1002/cpe.3416}
}

@inproceedings{nivre2020,
  author    = {Nivre, Joakim and de Marneffe, Marie-Catherine and Ginter, Filip and Haji\v{c}, Jan and Manning, Christopher D. and Pyysalo, Sampo and Schuster, Sebastian and Tyers, Francis and Zeman, Daniel},
  title     = {{Universal Dependencies} v2: An evergrowing multilingual treebank collection},
  booktitle = {Proceedings of the Twelfth International Conference on Language Resources and Evaluation ({LREC} 2020)},
  year      = {2020},
  publisher = {European Language Resources Association},
  address   = {Marseille, France},
  pages     = {4034--4043},
  url       = {https://aclanthology.org/2020.lrec-1.497/}
}

\end{document}